\documentclass[reprint,twocolumn,superscriptaddress, amssymb,aps,prl]{revtex4-1}

\usepackage{graphicx}
\usepackage{xcolor} 
\usepackage{amsmath}
\usepackage{hyperref}
\newcommand{\bra}{\left \langle}
\newcommand{\ket}{\right \rangle}

\begin{document}
\title{On-Demand Entanglement of Molecules in a \\ Reconfigurable Optical Tweezer Array}
\author{Connor M. Holland}
\email{These authors contributed equally to this work.}
\affiliation{Department of Physics, Princeton University, Princeton, New Jersey 08544 USA}
\author{Yukai Lu}
\email{These authors contributed equally to this work.}
\affiliation{Department of Physics, Princeton University, Princeton, New Jersey 08544 USA}
\affiliation{Department of Electrical and Computer Engineering, Princeton University, Princeton, New Jersey 08544 USA}
\author{Lawrence W. Cheuk}
\email{lcheuk@princeton.edu}
\affiliation{Department of Physics, Princeton University, Princeton, New Jersey 08544 USA}

\date{\today}

\begin{abstract}
Entanglement is crucial to many quantum applications including quantum information processing, simulation of quantum many-body systems, and quantum-enhanced sensing~\cite{nielsen2000chuang,Gisin2002,Preskill2018NISQ,amico2008entanglement,Pezze2018RMPMetrology}. Molecules, because of their rich internal structure and interactions, have been proposed as a promising platform for quantum science~\cite{demille2002quantum, Carr2009review,Blackmore2018QuantumReview}. Deterministic entanglement of individually controlled molecules has nevertheless been a long-standing experimental challenge. Here we demonstrate, for the first time, on-demand entanglement of individually prepared molecules. Using the electric dipolar interaction between pairs of molecules prepared using a reconfigurable optical tweezer array, we realize an entangling two-qubit gate, and use it to deterministically create Bell pairs~\cite{Bell1964}. Our results demonstrate the key building blocks needed for quantum information processing, simulation of quantum spin models, and quantum-enhanced sensing. They also open up new possibilities such as using trapped molecules for quantum-enhanced fundamental physics tests~\cite{augenbraun2020lasercoolingPrecMeas} and exploring collisions and chemical reactions with entangled matter. 
\end{abstract}
\maketitle
\maketitle

\section{Introduction}
Entanglement lies at the heart of quantum mechanics. It is both central to the practical advantage provided by quantum devices~\cite{nielsen2000chuang,Gisin2002,Preskill2018NISQ}, and important to understanding the behavior of many-body quantum systems~\cite{amico2008entanglement}. The ability to create entanglement controllably has been a long-standing experimental challenge. Molecules have been proposed as a promising platform for quantum simulation and quantum information processing because of their rich internal structure and long-lived interacting states~\cite{demille2002quantum, Carr2009review,Blackmore2018QuantumReview}. In the past two decades, much progress has been made in producing and controlling molecules at ultracold temperatures, both through coherent assembly of ultracold alkali atoms~\cite{Ni2008groundstate} and direct laser-cooling~\cite{Barry2014SrFMOT}. Rapid advances have been made in recent years, including the creation of degenerate molecular gases~\cite{demarco2019degKrb,schindewolf2022NaKEvap}, the creation of molecular magneto-optical traps~\cite{Barry2014SrFMOT,Truppe2017Mot,Anderegg2017MOT,Collopy2018MOT}, high-fidelity detection of single molecules~\cite{Liu2018NaCs,Anderegg2019Tweezer,christakis2022probing} and laser-cooling of complex polyatomic molecules~\cite{mitra2020CaOCH3,Vilas2022MOTCaOH}. In addition, coherent dipolar interactions have also been observed in bialkali molecules trapped in optical lattices~\cite{yan2013observation,christakis2022probing}.

\begin{figure}
	{\includegraphics[width=\columnwidth]{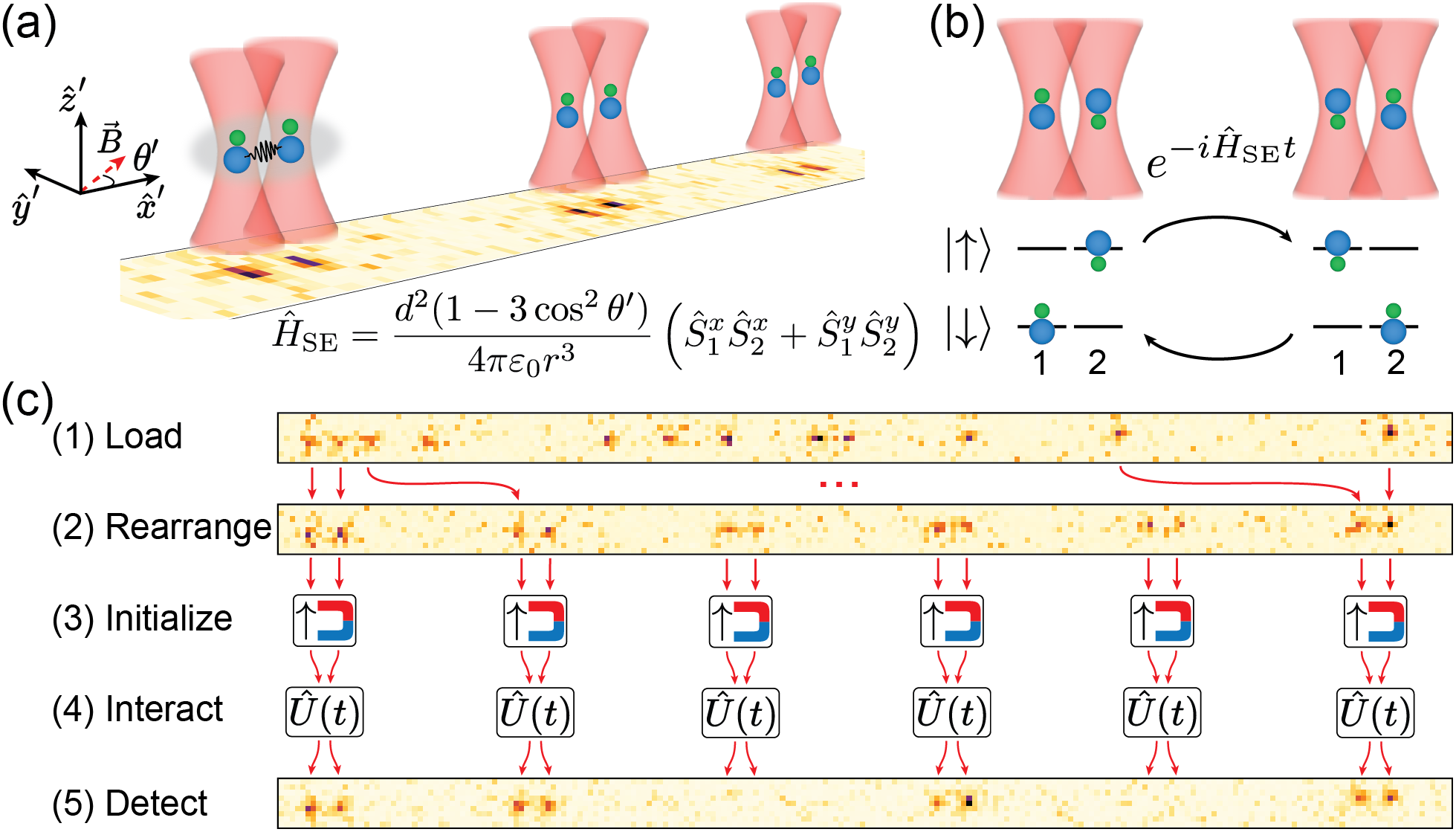}}
	\caption{\label{fig_1} Laser-cooled Molecules in a Reconfigurable Optical Tweezer Array. (a) Single CaF molecules trapped in an optical tweezer array are prepared into closely separated tweezer pairs. Molecules in each pair held by separate tweezer traps interact via the long-range electric dipolar interaction $\hat{H}_{\text{SE}}$. (b) The electric dipolar interaction leads to dipolar spin-exchange of rotational excitations. (c) Molecules are loaded stochastically, detected non-destructively, and rearranged into the desired 1D configuration. The molecules are then initialized into a single internal state and the pair separations are reduced in order to switch on interactions. After specific interaction times, the pairs are then separated and detected state-selectively.}
	\vspace{-0.2in}
\end{figure}

A major outstanding challenge to fully realizing the potential of molecules has been achieving deterministic entanglement with microscopic control. In this work, we report the first realization of a two-qubit entangling gate between individual laser-cooled molecules trapped in a reconfigurable optical tweezer array (Fig.~\ref{fig_1}a). The approach of molecular tweezer arrays~\cite{Liu2018NaCs,Anderegg2019Tweezer,Zhang2021NaCsArray,holland2022bichromatic} combines the microscopic controllability offered by reconfigurable optical tweezer traps~\cite{Endres2016atomarray,barredo2016atom,Cooper2018AEA,Norcia2018AEA,Saskin2019AEA} with the ability to generate entanglement through the electric dipolar interaction between molecules. We specifically make use of effective spin-exchange interactions that arise between rotational states to realize an iSWAP gate, which along with single qubit rotations, is sufficient for universal quantum computation~\cite{Ni2018gate} (Fig.~\ref{fig_1}b). We subsequently use this to entangle pairs of molecules into Bell states, which are prototypical entangled states~\cite{Bell1964} often used to benchmark quantum platforms such as photons~\cite{aspect1982experimental}, trapped ions~\cite{turchette1998deterministic}, neutral atoms~\cite{hagley1997generation}, superconducting circuits~\cite{steffen2006measurement}, nitrogen-vacancy centers in diamond~\cite{Neumann2008NV}, and semiconductor quantum dots~\cite{Zajac2018QDEntanglement}.

\section{Preparing Pairs of Laser-Cooled Molecules and Initializing Their Internal States}
Our work starts with single laser-cooled CaF molecules trapped in a dynamically reconfigurable array of optical tweezer traps~\cite{Anderegg2019Tweezer,holland2022bichromatic}. Through a series of steps involving laser-cooling, optical trapping and transport, single molecules are transferred from a magneto-optical trap into a 1D array of 37 identical optical tweezer traps with a uniform spacing of $4.20(6)\,\mu\text{m}$. Since our laser-cooling scheme relies on a closed optical cycle present only for the $X^2\Sigma(v=0, N=1)$ in CaF~\cite{stuhl2008Cycling}, the molecules initially loaded into the tweezers occupy a single rovibrational manifold.

\begin{figure}
	{\includegraphics[width=\columnwidth]{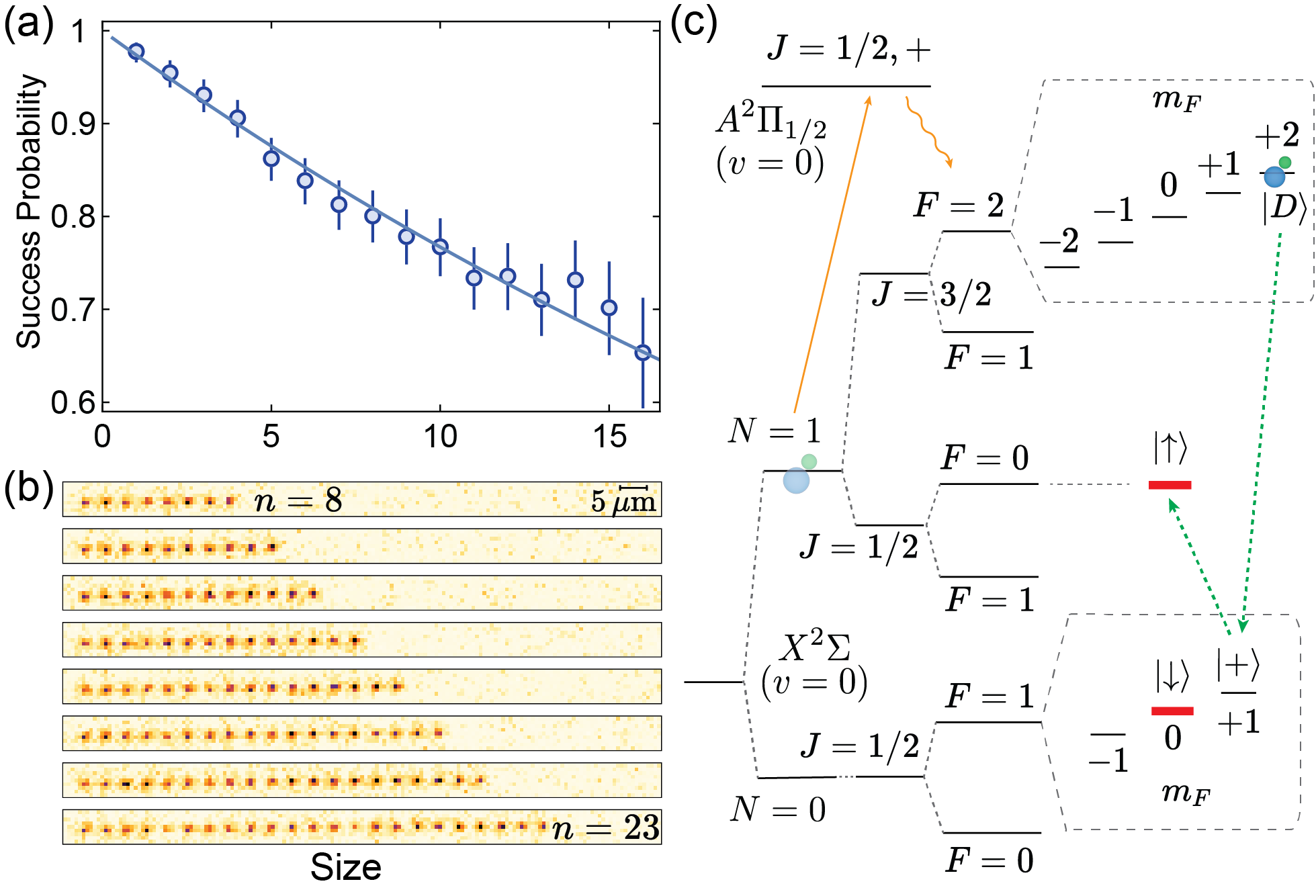}}
	\caption{\label{fig_2} Tweezer Rearrangement and Internal State Initialization. (a) Probability of creating defect-free molecular arrays via rearrangement. A fit to $p^n$, where $n$ is the array size, gives a single particle rearrangement fidelity of $p=0.974(1)$. (b) Exemplary images of defect-free arrays. (c) Optical pumping (orange arrow) prepares molecules in $|{D}\rangle$. Microwave sweeps (dashed green arrows) transfer $|{D}\rangle$ molecules to $|{\uparrow}\rangle$. 	}
	\vspace{-0.2in}
\end{figure}

To remove the randomness in tweezer occupation, we use a rearrangement approach pioneered in neutral atom experiments~\cite{Endres2016atomarray,barredo2016atom}. We non-destructively detect the tweezer occupations using a variant of $\Lambda$-imaging~\cite{Cheuk2018Lambda}. The empty tweezers are identified, switched off, and the remaining occupied tweezers are then rearranged into the desired 1D pattern. We characterize the rearrangement procedure by measuring the probability of successfully creating uniform arrays, and find a single particle rearrangement fidelity of 97.4(1)\%. As shown in Fig.~\ref{fig_2}a, we are able to create uniform arrays up to a size of $16$ with a probability $>0.6$. The rearrangement fidelity is limited by the non-destructive detection fidelity, with minimal loss (0.2(10)\%) due to movement of the tweezer traps.

After rearrangement, we initialize the internal state of the molecules, which are distributed among the 12 hyperfine states in the $X^2\Sigma(v=0, N=1)$ rovibrational manifold. To prepare molecules into a single hyperfine state, we optically pump molecules into $\left|D\ket=X^2\Sigma(v=0, N=1, J=3/2, F=2, m_F=2)$. Subsequent microwave sweeps along with an optical clean-out pulse transfer the molecules into the target final state $\left|\uparrow\ket = X^2\Sigma(v=0,N=1,J=1/2,F=0,m_F=0)$ (Fig. \ref{fig_2}c). The overall fidelity of preparing molecules in $|{\uparrow}\rangle$ is $82.4(11)\%$. Our preparation sequence ensures that the dominant preparation error is in the form of unoccupied tweezers, with a small contribution ($\approx1\%$) coming from molecules prepared in the incorrect internal state $\left|+\ket=X^2\Sigma(v=0, N=0,J=1/2, F=1,m_F=1)$. The state initialization errors come from imperfect microwave transfer, polarization impurity of the optical pumping light, and loss due to heating in the tweezer traps.

\section{Probing Single-Molecule Coherence}
In order to produce entanglement via the dipolar interactions between molecules, we require long coherence times compared to the typical interaction timescales of $h/J \sim 10\,\text{ms}$ at our tweezer separations. Achieving long coherence times for optically trapped molecules has been an ongoing experimental challenge with steady advances being made. For molecules, different internal states can experience different trapping potentials, which, in combination with motion due to finite temperature, can lead to decoherence. For $^1\Sigma$ bialkali molecules, long coherence times of different nuclear spin states have been reported~\cite{park2017second,gregory2021robust}, and work using ``magic'' trapping conditions have demonstrated extended coherence times between rotational states in both $^1\Sigma$ and $^2\Sigma$ molecules~\cite{neyenhuis2012anisotropic,seesselberg2018extending,lin2021anisotropic,burchesky2021rotcoh,christakis2022probing}. 

Since the effective spin-exchange interactions couple different rotational states, we desire long rotational coherence times between the two interacting states $\left| \uparrow \ket$ and $\left| \downarrow\ket=X^2\Sigma(v=0,N=0,J=1/2,F=1,m_F=0)$. Building upon previous work in CaF~\cite{burchesky2021rotcoh}, we identify a pseudo-magic trapping condition where both spin states experience approximately identical trapping potentials. Our pseudo-magic condition takes into account vector and tensor shifts and is achieved by applying a magnetic field orthogonal to the tweezer light polarization at a reduced tweezer depth compared to that used for initial loading and imaging. 
\begin{figure*}
	{\includegraphics[width=2\columnwidth]{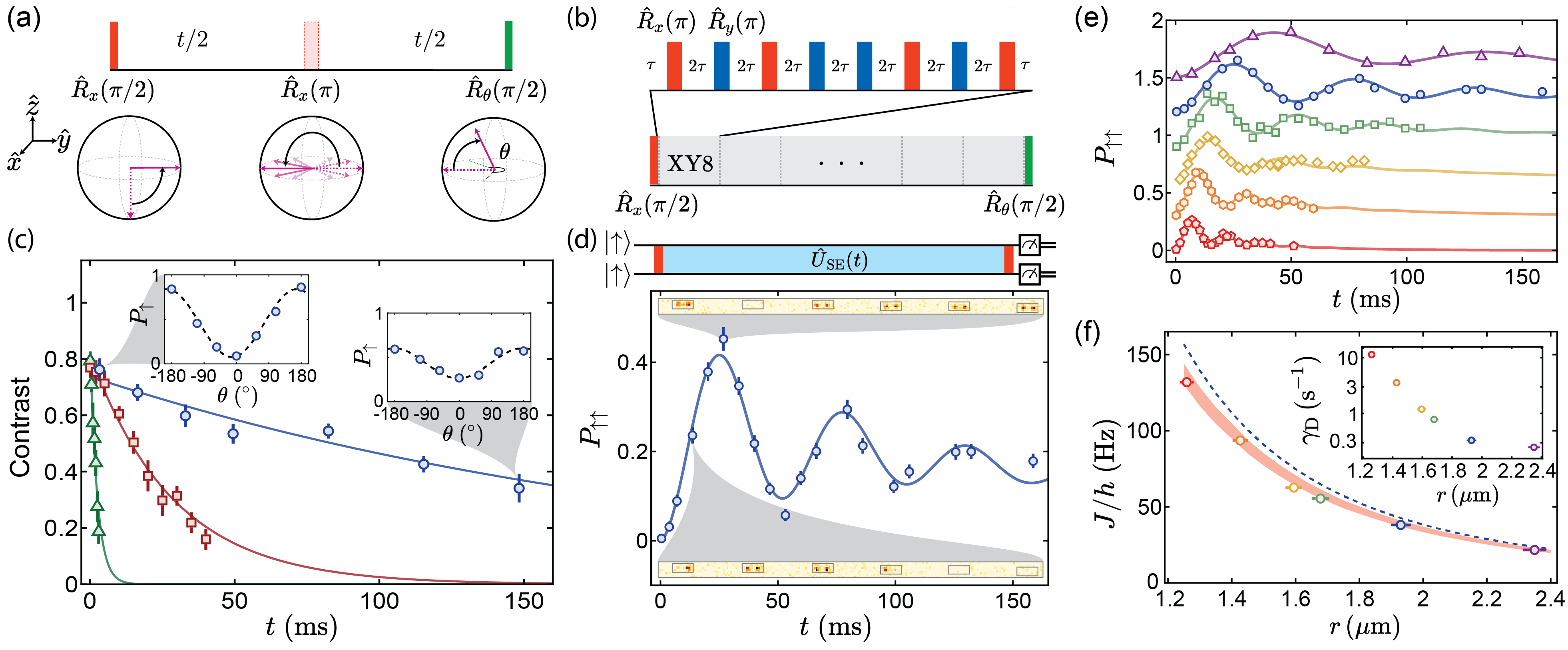}}
	\caption{\label{fig_3} Single Particle Coherence and Spin-Exchange Oscillations. (a) Ramsey pulse sequence used to measure rotational coherence. The bottom Bloch sphere diagrams show the action of the various pulses for a molecule initialized in $\left|\downarrow\ket$. (b) The XY8 dynamical decoupling sequence. (c) Ramsey contrast of non-interacting molecules versus free evolution time $t$. Green triangles, red squares, and blue circles show the cases when no spin-echo, one spin-echo, and the XY8 sequence is applied, respectively. Exponential fits give coherence times ($1/e$) of 2.5(3)\,ms, 29(2)\,ms, and 215(30)\,ms, respectively. Insets show exemplary Ramsey fringes with corresponding sinusoidal fits shown by the dashed lines. (d) Spin-exchange oscillations at a tweezer separation of $1.93(3)~\mu\text{m}$. Shown are the $\left|\uparrow\uparrow\ket$ populations measured after the Ramsey pulse sequence, $P_{\uparrow\uparrow}$, as a function of interaction time $t$, for molecular pairs initialized in $\left|\uparrow\uparrow\ket$. The solid curve is a fit to a phenomenological model. Insets show exemplary fluorescence images at the indicated times. (e) Spin-exchange oscillations at separations of $1.26(2)\,\mu\text{m}$ (red pentagons), $1.43(2)\,\mu\text{m}$ (orange hexagons), $1.60(2)\,\mu\text{m}$ (yellow diamonds), $1.68(2)\,\mu\text{m}$ (green squares), $1.93(3)\,\mu\text{m}$ (blue circles), and $2.35(3)\,\mu\text{m}$ (purple triangles). Curves are offset vertically by 0.3 for clarity. (f) The extracted spin-exchange strength $J$ versus pair separation $r$. The light red band indicates the theoretical prediction taking into account the finite temperature of the molecules and the uncertainty in the electric dipole moment of CaF. The dashed blue curve shows the prediction without taking into account finite temperature. Inset shows the single particle loss rate $\gamma_\text{D}$ versus pair separation $r$.}
	\vspace{-0.2in}
\end{figure*}

To measure the resulting coherence time, we prepare tweezer pairs with only one molecule initialized in $\left|\uparrow\ket$. We next apply a Ramsey pulse sequence consisting of two $\pi/2$ microwave pulses (first pulse along $\hat{x}$, second pulse along $\hat{n} = \cos \theta \, \hat{x} + \sin\theta\, \hat{y}$) separated by a variable free evolution time (Fig.~\ref{fig_3}a). The remaining fraction of $\left|\uparrow\ket$ molecules, $P_{\uparrow}$, oscillates as a function of $\theta$, with the oscillation amplitude directly measuring the coherence. Fitting to an exponential decay curve yields a bare coherence time $T_2^*$ of $2.5(3)\,\text{ms}$. Adding a spin-echo improves the coherence time to $T_2= 29(2)\,\text{ms}$. Following previous work exploring dipolar interactions of KRb molecules in an optical lattice~\cite{yan2013observation,li2022tunable}, we implement the XY8 dynamical decoupling sequence depicted in Fig.~\ref{fig_3}b, and find that the $1/e$ coherence time is further extended to 215(30)\,ms (Fig.~\ref{fig_3}c). This is consistent with our understanding that the bare coherence times are primarily limited by slow (ms-timescale) fluctuations of ambient magnetic fields.

\section{Observing Coherent Dipolar Spin-Exchange Interactions} 
Having achieved sufficiently long rotational coherence times, we next set out to observe coherent spin-exchange interactions. The long-range electric dipolar interaction between the molecules gives rise to resonant exchange of rotational excitations between $\left|\uparrow\ket$ and $\left|\downarrow\ket$. The resulting spin-exchange interaction is described by the Hamiltonian
\begin{equation}
\hat{H}_{\text{SE}} = \frac{J}{2} \Big(\hat{S}_{1}^+ \hat{S}_{2}^- + \hat{S}_{1}^- \hat{S}_{2}^+\Big)=J\left(\hat{S}_{1}^x \hat{S}_{2}^x + \hat{S}_{1}^y \hat{S}_{2}^y\right),
\nonumber
\end{equation}
where $\hat{S}_{i}^+$, $\hat{S}_{i}^-$, $\hat{S}_{i}^x$, $\hat{S}_{i}^y$ are spin-1/2 operators for molecule $i$ and 
\begin{equation}
J =\frac{d^2}{4\pi \varepsilon_0} \frac{1}{r^3} (1-3\cos^2\theta'),
\nonumber
\end{equation}
with $d = \langle{\uparrow}|\hat{\mathbf{d}}|{\downarrow}\rangle$ being the transition dipole moment, $r=|\vec{r}|$ being the intermolecular separation, $\theta'$ being the angle between $\vec{r}$ and the quantization axis, and $\varepsilon_0$ being the free space permittivity. Starting with two molecules in a product state, time evolution under $\hat{H}_{\text{SE}}$ can lead to entanglement. For example, two molecules initially prepared in the product state $\left|\uparrow\ket \otimes \left| \downarrow\ket$ become maximally entangled after interacting for a time $t=\pi \hbar /(2J)$. 

To observe the effect of spin-exchange interactions, we first create pairs of $\left|{\uparrow}\ket$ molecules at an initial separation of $4.20(6)\,\mu\text{m}$, over which interactions are negligible. We next reduce the pair separation to $1.93(3)\,\mu\text{m}$ over 3\,ms, where the interaction strength $J$ ($J= h \times 43\,\text{Hz}$) becomes appreciable on the coherence timescale. Subsequently, we apply the Ramsey pulse sequence used above with $\theta=0$. To retain long coherence times, the XY8 decoupling pulses are kept on during the free evolution time. Because the $\pi$-pulses in the XY8 sequence leave $\hat{H}_{\text{SE}}$ unchanged, spin-exchange interactions are preserved~\cite{choi2020robust}.

For a molecular pair initialized in $\left|{\uparrow}{\uparrow}\ket$, the resulting state following the Ramsey sequence is given by 
\begin{equation}
\left|\psi\ket = i e^{-i \frac{Jt}{4\hbar}} \left(\sin \left(\frac{Jt}{4\hbar}\right)\left|\uparrow \uparrow \ket + i\cos\left(\frac{Jt}{4\hbar}\right)\left|\downarrow\downarrow\ket \right),
\end{equation}
and $P_{\uparrow\uparrow}$ oscillates at an angular frequency of $J/(2\hbar)$. As shown in Fig.~\ref{fig_3}d, we indeed observe oscillations of $P_{\uparrow\uparrow}$, directly revealing the presence of coherent spin-exchange interactions. By varying the pair separation between $1.26(2)~\mu\text{m}$ and $2.35(3)~\mu\text{m}$, we verify that the interaction strength $J$, as extracted from the oscillation frequency, approximately scales as $1/r^3$, as expected for dipolar interactions (Fig.~\ref{fig_3}f). 

We observe that the oscillations decay faster when the molecules are closer. This decay could be due to molecular loss or reduced single particle coherence times at close separations. We verify that even at the closest spacing of $1.26(2)\,\mu\text{m}$, tunneling of molecules between adjacent tweezers, which could lead to loss, is negligible. Separately measured single particle loss rates show that molecular loss increases at short separations and significantly contributes to damping (Fig~\ref{fig_3}f inset). We believe the molecular loss is likely due to parametric heating specific to our scheme of generating tweezer traps using an acousto-optical deflector (AOD), and can be circumvented with other tweezer generation techniques. Comparison with independently measured single particle decoherence rates point to additional damping that could be due to thermal motion of molecules in the tweezer traps. 

Our work is the first observation of coherent interactions between individually prepared molecules, and also between laser-cooled molecules. We note in passing that experiments with bulk samples of bialkali molecules in optical lattices have observed coherent spin-exchange both through macroscopic measurements~\cite{yan2013observation,li2022tunable} and also recently with single molecule resolution~\cite{christakis2022probing}.

\section{Realizing a Two-Qubit Gate and Creating Bell Pairs}
We next create a maximally entangling two-qubit gate, which, together with single qubit rotations, is sufficient for universal quantum computing. As proposed in~\cite{Ni2018gate}, we realize an iSWAP gate between two molecules by allowing them to interact via $\hat{H}_{\text{SE}}$ for a specific time of $T=\pi\hbar/J$. We use this gate to generate maximally entangled Bell pairs by applying two additional $\pi/2$ pulses along the $x$-axis before and after the gate. This sequence converts a molecular pair prepared in $\left|\uparrow\uparrow\ket$ into the Bell state $\left|\psi_B\ket =\frac{1}{\sqrt{2}}(\left |\uparrow\uparrow\ket + i \left |\downarrow\downarrow\ket)$. As a compromise between maximizing $J$ and minimizing heating loss, we choose a pair separation of $1.93(3)\,\mu\text{m}$ for creating $\left|\psi_B\ket$.

\begin{figure*}
	{\includegraphics[width=2\columnwidth]{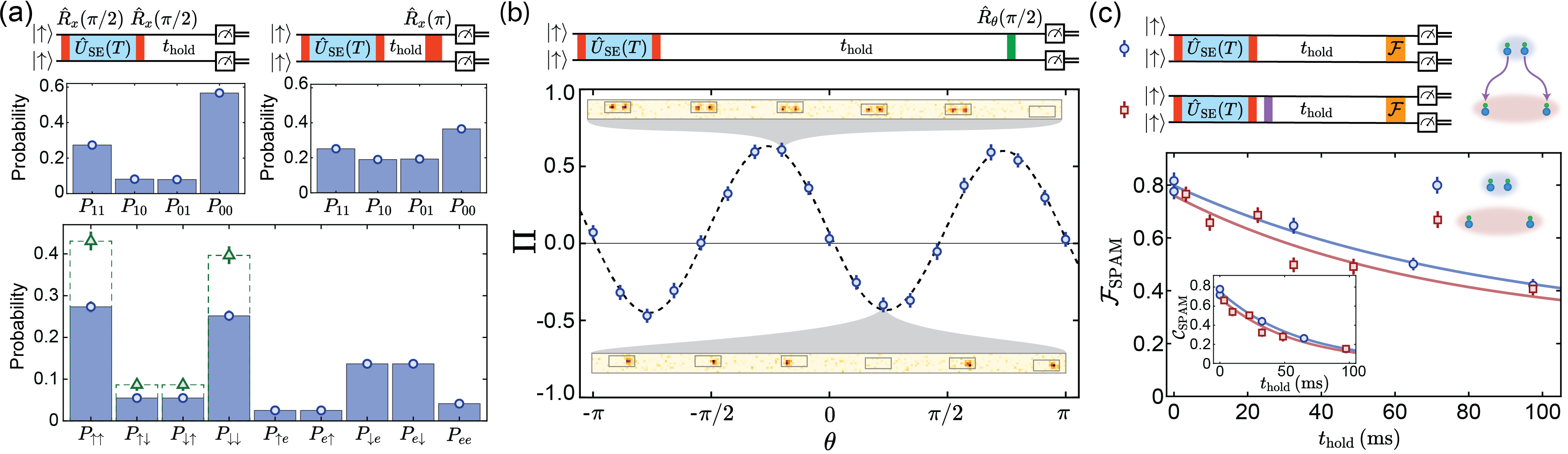}}
	\caption{\label{fig_4} Creating and Probing Bell Pairs. (a) Top left panel shows the two-tweezer detection probabilities $P_{ij}$ for the Bell pairs. Right panel shows corresponding probabilities when an additional $\pi$ pulse is applied before measurement. Bottom panel shows the full probabilities extracted from the top panels. The SPAM-corrected populations for the $\left|\uparrow\ket - \left|\downarrow\ket$ subsystem are shown by the green triangles. (b) Probing Bell state coherence through parity oscillations. The coherence $\mathcal{C}$ between $|{\uparrow\uparrow}\rangle$ and $|{\downarrow\downarrow}\rangle$ is obtained from the amplitude of the oscillations in $\Pi$ as a function of $\theta$. Insets show exemplary images at the maxima and minima of $\Pi$. (c) SPAM-corrected Bell state fidelity $\mathcal{F}_\text{SPAM}$ versus hold time $t$ at the initial pair creation distance of 1.93(3)\,$\mu\text{m}$ are shown in blue circles. Red squares show the corresponding data for pairs separated to a larger distance of $4.20(6)~\mu\text{m}$ following creation. Inset shows the contrast $\mathcal{C}_\text{SPAM}$ versus hold time $t$ for the two separations. The extracted $1/e$ lifetimes are ($\mathcal{F}_\text{SPAM}$, $\mathcal{C}_\text{SPAM}$)= ($85(5)\,\text{ms}$, $61(3)\,\text{ms}$) at the Bell creation distance, and ($70(16)\,\text{ms}$, $57(8)\,\text{ms}$) at the larger separation.}
	\vspace{-0.2in}
\end{figure*}

To measure the fidelity $\mathcal{F} =\bra \psi_B \right| \rho \left|\psi_B\ket $ of creating $\left|\psi_B\ket$, where $\rho$ is the density matrix, we make use of the relation~\cite{turchette1998deterministic}
\begin{equation}
 \mathcal{F} = \frac{1}{2} (P_{\uparrow\uparrow} + P_{\downarrow\downarrow} + \mathcal{C}),
\end{equation}
where $P_{\uparrow\uparrow}$ ($P_{\downarrow\downarrow}$) are the probabilities of measuring $\left| \uparrow \uparrow\ket$ $(\left|\downarrow \downarrow\ket)$, and $\mathcal{C}$ is the amplitude of the coherence between $\left |\uparrow\uparrow\ket$ and $\left|\downarrow\downarrow\ket$. 

Our imaging scheme only detects $\left|\uparrow\ket$ molecules. To measure $P_{\uparrow\uparrow}$, we separate the two molecules after preparing $\left|\psi_B\ket$ and subsequently measure the probability that both tweezers in a pair appear bright. To measure $P_{\downarrow\downarrow}$, we apply an additional $\pi$-pulse prior to detection (Fig.~\ref{fig_4}a) to convert molecules from $\left|\downarrow\ket$ to $\left|\uparrow\ket$ for detection. To obtain the coherence envelope $\mathcal{C}$, we apply a $\pi/2$ pulse about a variable axis $\hat{n} =\cos \theta\,\hat{x} + \sin \theta\,\hat{y}$ after preparing $\left|\psi_B\ket$. From the two-tweezer probabilities $P_{ij}$, we construct the parity signal $\Pi = P_{11}+ P_{00}-P_{10}- P_{01}$, where $1(0)$ denotes a bright (dark) tweezer site. 

For a general density matrix that includes the possibility of empty tweezers ($\left|e\ket$) due to imperfect state preparation, $\Pi$ can display modulation at $2\theta$ and $\theta$. The $2\theta$ modulation is directly related to the coherence between $\left|\uparrow\uparrow\ket$ and $\left|\downarrow\downarrow\ket$, while the $\theta$ modulation is related single molecule coherences, such as that between $\left|e\uparrow\ket$ and $\left|e\downarrow\ket$. The amplitude of the $2\theta$ modulation directly gives $\mathcal{C}$. As shown in Fig.~\ref{fig_4}b, we find that $\Pi$ only oscillates at $2\theta$. The lack of oscillations at $\theta$ is consistent with the pulse sequence used. Single molecules prepared in $\left|e\uparrow\ket$ or $\left|\uparrow e\ket$ effectively experience a $\pi$-pulse and no single particle coherence is created. Correcting only for measurement errors, we obtain a Bell state fidelity $\mathcal{F} = 0.54(1)$. The state-preparation and measurement (SPAM) corrected fidelity is $\mathcal{F}_\text{SPAM}=0.80(2)$. We believe the fidelity to be limited in part by the single molecule decoherence time and the finite temperature of the molecules. We leave the investigation and improvement of these effects to future work.

We note in passing that the modulation of $\Pi$ at $2\theta$ directly shows the enhanced metrological sensitivity of the Bell state, which is a two-particle realization of a Greenberger-Horne-Zeilinger state (GHZ) state, a class of entangled states known to reach the Heisenberg limit in quantum metrology~\cite{Pezze2018RMPMetrology}. Specifically, the $2\theta$ modulation in $\Pi$ directly illustrates the two-fold increase in sensitivity of the Bell state compared to unentangled states. 

In addition, the phase of the $2\theta$ oscillation in $\Pi$ measures the relative phase between $\left| \uparrow \uparrow\ket$ and $\left| \downarrow \downarrow\ket$ in the Bell state, which is sensitive to the sign of $J$. As expected, the observed phase of the oscillation shows that $J>0$, indicating anti-ferromagnetic spin-exchange interactions. In the context of quantum simulation, our measurements demonstrate the fundamental building block for simulating anti-ferromagnetic XY spin models.

Having demonstrated deterministic creation of Bell pairs, we next probe their lifetime. We measure the Bell state fidelity $\mathcal{F}_\text{SPAM}$ as a function of hold time and find a lifetime of $85(5)\,\text{ms}$, largely consistent with but somewhat shorter than that expected from uncorrelated single particle decoherence. We also examine whether the Bell pairs survive when separated to larger distances following their creation. Specifically, we separate the Bell pairs to a distance of $4.20(6)~\mu\text{m}$ over $3\,\text{ms}$ following their creation, and measure $\mathcal{F}_\text{SPAM}$ as a function of hold time. Within experimental uncertainty, we find a lifetime identical to the case without separation (Fig.~\ref{fig_4}c). This ability to preserve entanglement while separating molecules could allow one to bypass the limited range of dipolar interactions through movement of molecules, and obtain arbitrary connectivity useful for quantum simulation and information processing. Similar abilities to preserve entanglement have recently been demonstrated in atomic tweezer arrays~\cite{bluvstein2022quantum}.

\section{Outlook}
In summary, we have demonstrated on-demand creation of entangled Bell pairs of molecules in a reconfigurable optical tweezer array. Our work is enabled by several advances we have made in controlling laser-cooled molecules, including initialization of defect-free molecular arrays, achievement of long rotational coherence times in tightly-focused tweezer traps, and observation of coherent dipolar interactions between individual laser-cooled molecules. Our results demonstrate key building blocks for simulating quantum spin models, processing quantum information, and realizing quantum-enhanced metrology in the emerging platform of molecular tweezer arrays. Specifically for quantum simulation and information processing, the interactions between long-lived molecular states could offer long evolution times and deep circuit depths competitive with other platforms such as Rydberg atom arrays, trapped ions, and superconducting qubits. Our work on entangling molecules on-demand, combined with the rapid progress in extending laser-cooling to molecular species of increasing complexity~\cite{mitra2020CaOCH3,Vilas2022MOTCaOH}, opens new research avenues such as quantum-enhanced precision measurement using trapped molecules~\cite{augenbraun2020lasercoolingPrecMeas} and explorations of molecular collisions~\cite{Cheuk2020collisions} and chemical reactions with entangled matter.

\section{Acknowledgements}
We thank Waseem Bakr's group and Sarang Gopalakrishnan for fruitful discussions. We also thank Waseem Bakr, Jeff Thompson, Samuel J. Li, and Christie Chiu for careful readings of the manuscript. This work is supported by the National Science Foundation under Grant No. 2207518. L.W.C. acknowledges support by the Sloan Foundation.

\end{document}


\title{Supplementary Material for ``On-Demand Entanglement of Molecules in a Reconfigurable Optical Tweezer Array"}
\author{Connor M. Holland}
\email{These authors contributed equally to this work.}
\affiliation{Department of Physics, Princeton University, Princeton, New Jersey 08544 USA}
\author{Yukai Lu}
\email{These authors contributed equally to this work.}
\affiliation{Department of Physics, Princeton University, Princeton, New Jersey 08544 USA}
\affiliation{Department of Electrical and Computer Engineering, Princeton University, Princeton, New Jersey 08544 USA}
\author{Lawrence W. Cheuk}
\email{lcheuk@princeton.edu}
\affiliation{Department of Physics, Princeton University, Princeton, New Jersey 08544 USA}

\date{\today}
\maketitle

\subsection{Apparatus Overview}
\begin{figure}
	{\includegraphics[width=\columnwidth]{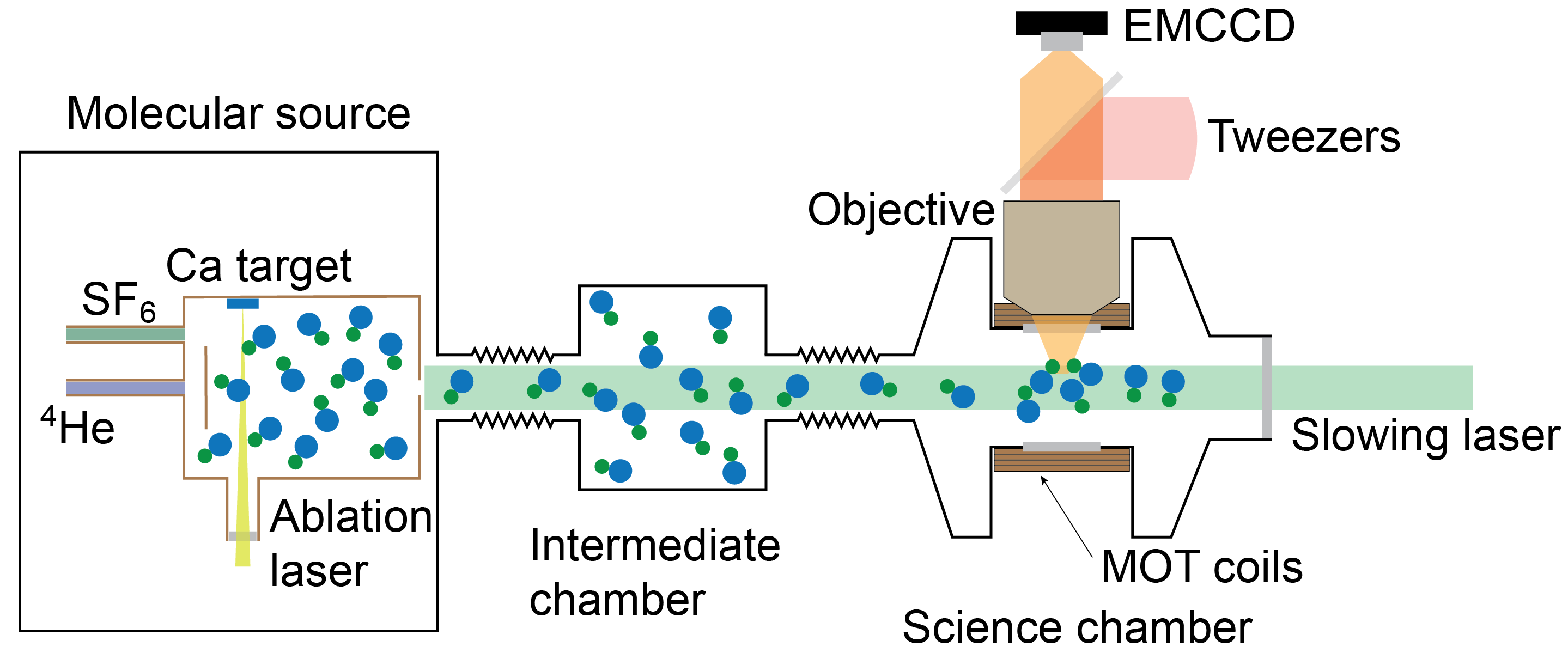}}
	\caption{\label{fig:apparatus} Schematic overview of apparatus.	}
	\vspace{-0.2in}
\end{figure}

Our apparatus consists of a molecular source chamber, intermediate chamber, and science chamber connected in series (Fig.~\ref{fig:apparatus}). CaF molecules are produced in the molecular source chamber, after which they are slowed and loaded into a magneto-optical trap (MOT) in the center of the science chamber~\cite{Truppe2017Mot}. Molecules are then optically trapped and transferred into the optical tweezer traps, which all occur in the science chamber. A microscope objective mounted in a recessed viewport allows creation of tightly focused tweezer beams and high resolution imaging of single molecules. Details about the apparatus can be found in~\cite{Lu2021Ring,holland2022bichromatic}.

\subsection{Preparing Single Molecules In Optical Tweezer Traps}
Detailed descriptions of how we prepare CaF molecules into an optical dipole trap and into an optical tweezer array are contained in~\cite{Lu2021Ring} and~\cite{holland2022bichromatic}.

In short, we first create a cryogenic buffer gas beam of CaF molecules via laser ablation of a Ca target in the presence of $\text{SF}_6$~\cite{Hutzler2012CBGB}. The molecular beam has significant population in the laser-coolable $X^2\Sigma (v=0, N=1)$ manifold and is slowed via chirped laser slowing~\cite{Truppe2017chirp,Holland2021GSA}. The slowed molecules are captured into a DC magneto-optical trap (MOT) made using a quadrapole magnetic field along with beams addressing the $X^2\Sigma (v=0, N=1) - A^2\Pi_{1/2}(v=0, J=1/2,+)$ transition along three directions~\cite{Tarbutt2015DCMOT,Truppe2017Mot}. Vibrational repumping light addressing the $X^2\Sigma (v, N=1) - A^2\Pi_{1/2}(v-1, J=1/2,+)$ transitions for $v=1,2,3$ is simultaneously applied.

After the MOT is loaded, it is compressed via intensity and magnetic gradient ramps~\cite{Truppe2017Mot}. Subsequently, the MOT is switched off and the molecular sample is further compressed using a dynamically tuned ring-shaped repulsive trap in the presence of $\Lambda$-cooling~\cite{Cheuk2018Lambda,Lu2021Ring}. The molecules are then transferred into an attractive optical trap in the presence of $\Lambda$-cooling~\cite{Anderegg2018ODT,Cheuk2018Lambda}. The attractive trap is made using a focused and retro-reflected 1064\,nm beam with a Gaussian waist $w_0=60(7)\,\mu\text{m}$. The resulting lattice depth is $V\approx 600\,\mu\text{K}$. 

The molecules are then optically transported to the focal plane of the microscope objective, which is $\approx 1\,\text{mm}$ displaced from the center of the MOT. The transport is accomplished by moving the position of the 1D lattice over 30\,ms to its final location. Finally, the optical tweezer array is switched on in the presence of $\Lambda$-cooling. Over 60\,ms, molecules are cooled into a uniform array of 37 optical tweezer traps. The tweezer traps are stochastically loaded with a probability of 30 to 40\%. Following loading, the optical lattice is ramped down and single molecules are held only by the optical tweezer beams. Release-and-recapture measurements show molecular temperatures of $\approx 100\,\mu\text{K}$ at full tweezer depth.

\subsection{Optical Tweezer Trap Array Generation}
The 1D optical tweezer trap array is created by sending a single 781\,nm laser beam through an acousto-optical deflector (AOD) driven by multiple rf frequencies, and focusing the diffracted beams through a high numerical aperture microscope objective (NA = 0.65). The waveforms driving the AOD are generated via a computer-controlled arbitrary waveform generator (AWG), which are amplified by a high power rf amplifier (5W). The waveforms are optimized for minimal peak amplitude to reduce non-linearities due to saturation in the amplifier~\cite{Endres2016atomarray}. The intensities of each diffracted tweezer beam are balanced to 1\%  at an intermediate imaging plane. Differential ac Stark shift measurements show that this procedure balances the tweezer depths to within $\sim10\%$~\cite{holland2022bichromatic}. Each tweezer trap has a beam waist of $w_0 = 730\,\text{nm}$ and a maximum trap depth of $V= k_B \times 1.28\,\text{mK}$. During initial tweezer loading, the traps are configured to have a separation of $4.20(6)\,\mu\text{m}$ (AOD frequency spacing of 1\,MHz). Subsequent arrangements and movements of tweezer positions are done by sending in the appropriate waveforms to the AOD.

\subsection{Detection and Measurement Correction} 
\label{sec:meascorr}
We detect molecules in the optical tweezer traps using a variant of $\Lambda$-imaging~\cite{Cheuk2018Lambda,Anderegg2019Tweezer}, which provides fluorescence while simultaneously cooling the molecules. To perform imaging, we send in two retro-reflected beams with $\Lambda$-cooling light in the plane orthogonal to the objective (vertical) axis. To provide cooling along the vertical direction, a single beam oriented at $47^\circ$ with respect to the objective axis is sent in. Because of the tight trapping provided by the optical tweezer traps, this imbalanced 5-beam configuration still performs well. To detect the molecules, the resulting molecular fluorescence is collected through the objective and imaged onto an EMCCD camera. 

During imaging, only molecules in the $X^2\Sigma(v=0, N=1)$ rotational manifold emit light; molecules in $X^2\Sigma(v=0, N=0)$ all appear dark. In all data used in this work, two different imaging durations are used. For initial detection preceding and immediately following tweezer rearrangement, we use a 7\,ms long imaging pulse that is intended to be non-destructive. For final detection, we use a longer 30\,ms imaging pulse that is optimal for detection fidelity. 

\begin{figure}
	{\includegraphics[width=\columnwidth]{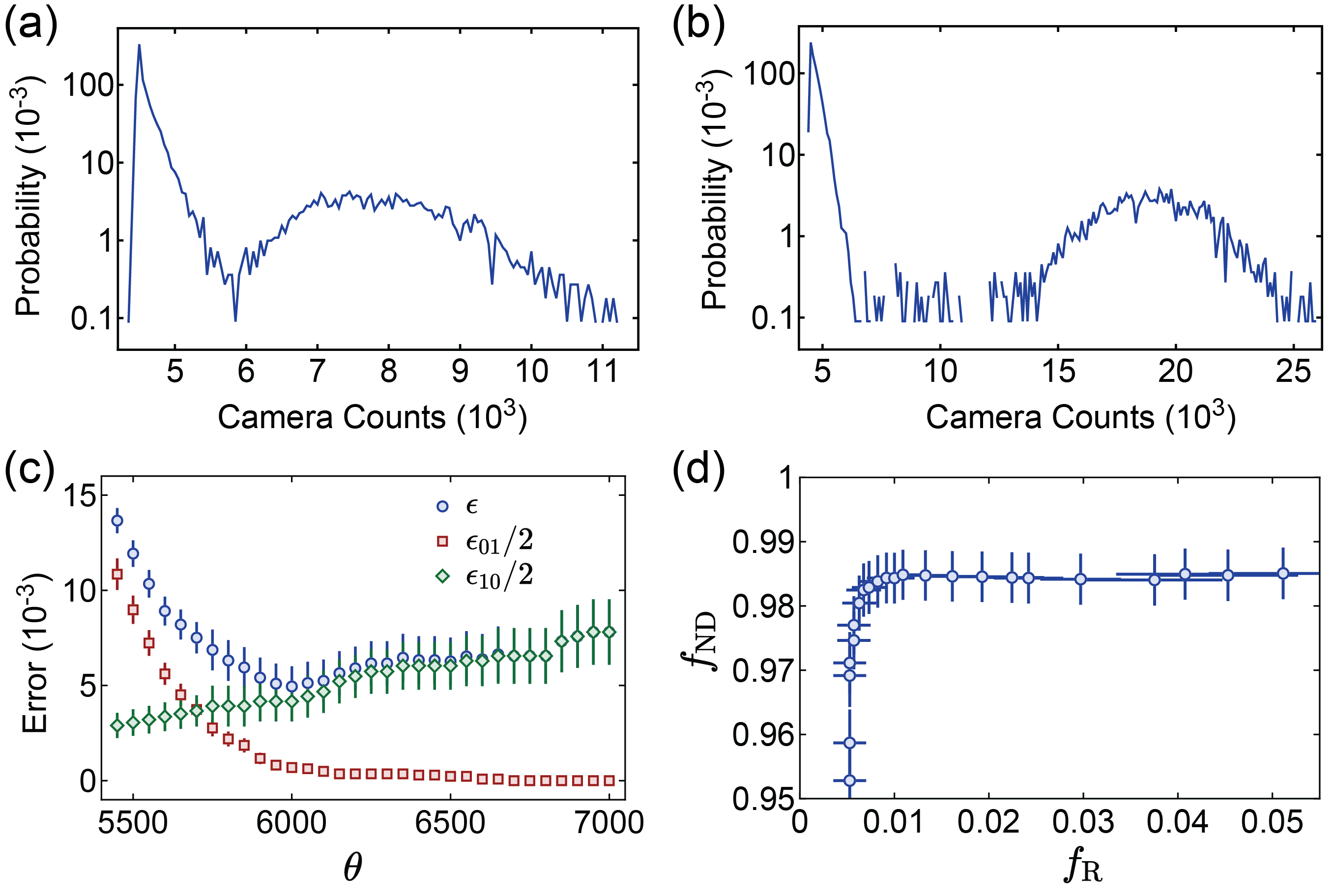}}
	\caption{\label{fig:histogramanderrors} Detection Performance. (a) Histogram of camera counts for the $7\,\text{ms}$ non-destructive imaging sequence. (b) Histogram of camera counts for the $30\,\text{ms}$ non-destructive imaging sequence. (c) As a function of the classification threshold $\theta$, the probability of misclassifying a tweezer instance at 50\% filling, $\epsilon$, shown in blue circles, has contributions $\epsilon_{10}/2$ (green diamonds) and $\epsilon_{01}/2$ (red squares). (d) The non-destructive fidelity $f_\text{ND}$ versus the data rejection rate $f_\text{R}$. Symbol definitions and our method to extract the quantities in (c) and (d) can be found in Ref. \cite{holland2022bichromatic}.}
	\vspace{-0.2in}
\end{figure}

The images are then analyzed to determine the occupation of each tweezer trap. We first sum camera counts over each image region corresponding to a tweezer trap. If the total counts exceed the detection threshold, the tweezer is classified as filled (1); otherwise, it is classified as empty (0). Using a procedure developed in previous work~\cite{holland2022bichromatic}, the error probabilities of misidentifying an occupied tweezer trap as empty ($\epsilon_{10}$) and vice versa ($\epsilon_{01}$) are determined. In  in Fig.~\ref{fig:histogramanderrors}, imaging histograms and the error probabilities as a function of detection threshold are shown. 

The error probabilities are then used to correct the measured data. Since the detected probabilities $P_{\text{det},i}$ are linear combinations of the true probabilities $P_i$~\cite{Schine2022gate}, one can write this in matrix form as $\mathbf{P_{det}} = \mathbf{M}\cdot \mathbf{P}$, where the entries of $\mathbf{M}$ are appropriate combinations of $\epsilon_{01}$ and $\epsilon_{10}$. Inverting the equation gives $\mathbf{P} = \mathbf{M}^{-1}\cdot \mathbf{P_{det}}$. 

For single particle probabilities $\mathbf{P} = \{P_1,P_0\}$, $\mathbf{M}_1$ is given by
\begin{equation}
\mathbf{M}_1 = \left(\begin{array}{cc}
\bar{\epsilon}_{10} & \epsilon_{01} \\
\epsilon_{10} & \bar{\epsilon}_{01} 
\end{array}\right),
\end{equation}
where $\bar{\epsilon} = 1-\epsilon$.

For two-particle probabilities $\mathbf{P} = \{P_{11},P_{10},P_{01},P_{00}\}$, $\mathbf{M}_2=\mathbf{M}_1\otimes\mathbf{M}_1$ is given by
\begin{equation}
\mathbf{M}_2 = \left(\begin{array}{cccc}
\bar{\epsilon}_{10}^2 & \bar{\epsilon}_{10}\epsilon_{01} &\epsilon_{01}  \bar{\epsilon}_{10}& \epsilon_{01}^2 \\
\bar{\epsilon}_{10}\epsilon_{10} & \bar{\epsilon}_{10}\bar{\epsilon}_{01} &\epsilon_{01} \epsilon_{10}& \epsilon_{01}\bar{\epsilon}_{01}  \\
\epsilon_{10} \bar{\epsilon}_{10}& \epsilon_{10}\epsilon_{01}&\bar{\epsilon}_{01}\bar{\epsilon}_{10} & \bar{\epsilon}_{01}\epsilon_{10}  \\
\epsilon_{10}^2 &\epsilon_{10}  \bar{\epsilon}_{01}&\bar{\epsilon}_{01}\epsilon_{10}& \bar{\epsilon}_{01}^2 
\end{array}\right)
\end{equation}

\subsection{Rearrangement of Molecular Arrays}
For the rearrangement procedure, we first determine the tweezer occupations using a $7\,\text{ms}$ imaging pulse following tweezer loading. The tweezers that are empty are switched off and the occupied ones are rearranged into one of two configurations: a uniformly-spaced defect-free array with $4.20(6)\,\mu\text{m}$ between tweezers or pairs of molecules spaced by $4.20(6)\,\mu\text{m}$ with $29.4(4)\,\mu\text{m}$ between the pairs. Following rearrangement, the molecules are cooled for 3\,ms to remove any heating that arises during tweezer movement. They are then imaged once more non-destructively ($7\,\text{ms}$). This image is used in the data analysis, and allows post-selection on successfully rearranged pairs.

To estimate the fidelity of the rearrangement process, we measure the success rate of creating a defect-free array of length $N$~\cite{Endres2016atomarray}. In detail, for each shot, we rearrange the molecules into a uniformly-spaced array with a length equal to the initially detected number of molecules. Next, we extract the raw success probabilities $P_{N,\text{raw}}$ of generating a defect-free array with length of at least $N$ starting from the left-most tweezer trap. Since a defect-free array can always be shortened deterministically with 100\% probability by turning off tweezer traps, $P_{N,\text{raw}}$ coincides with the success rate of generating a defect-free array of length $N$.

We correct for measurement errors in $P_{N,\text{raw}}$ using the procedure in the previous section to obtain $P_{N,m}$. At the classification threshold used for the final image, $\epsilon_{01}<0.01\%$ and $\epsilon_{10}=1.2(16)\%$. Since $\epsilon_{01}$ is negligible, the measurement-corrected success probability is obtained via
\begin{equation}
P_{N,m}=\frac{P_{N,\text{raw}}}{(1-\epsilon_{10})^N}
\end{equation}
Assuming the rearrangement errors to be equal at every site and uncorrelated, $P_{N,m}=p^N$, where $p$ is the single-particle rearrangement fidelity. The extracted rearrangement fidelity (Fig.~2) is found to be $p=97.4(1)\%$. Accounting for the nondestructive imaging fidelity $f_{\text{ND}} = 98.3(11)\%$ and the estimated survival of $99.3(1)\%$ during the 3\,ms cooling pulse, we find a rearrangement loss of $0.2(10)\%$, consistent with zero.

\subsection{Internal State Initialization}
For the coherent spin-exchange interaction data and the Bell state data, molecules are initialized in $|{\uparrow}\rangle =X^2\Sigma(v=0, N=1, J=1/2, F=0, m_F=0)$ using the scheme briefly described in the main text. In detail, we first optically pump the molecules into $\left|D\ket=X^2\Sigma(v=0, N=1, J=3/2, F=2, m_F=2)$. Next, a Landau-Zener microwave sweep transfers molecules from $|D\rangle$ to $\left|+\ket=X^2\Sigma(v=0, N=0,J=1/2, F=1,m_F=1)$. A $350\,\mu\text{s}$-long resonant light pulse then removes all molecules remaining in $X^2\Sigma (v=0,N=1)$. Finally, a Landau-Zener sweep transfers the molecules from $|+\rangle$ to $\left |\uparrow\ket = X^2\Sigma(v=0,N=1,J=1/2,F=0,m_F=0)$. The optical pumping is carried out at a tweezer separation of $4.20(6)\,\mu\text{m}$. The tweezers are moved into the desired pair separation afterwards. Note that our scheme ensures that the tweezers are empty, filled with $\left|+\ket$ molecules or filled with $\left|\uparrow\ket$ molecules.

From state-resolved measurements, we estimate that the purity of optical pumping into $\left|D\ket$ is $91.0(15)\%$, and that $5.6(8)\%$ of the molecules are lost between the beginning of non-destructive imaging and the end of optical pumping. Of this loss, 1.7(12)\% arises from non-destructive imaging, while the remaining 3.9(13)\% is attributed to recoil heating out of the trap during the optical pumping process (30 photons scattered in total), and heating due to intensity noise on the tweezer light. The microwave transfer efficiency from $\left|D\ket$ to $\left|+\ket$ is measured to be $97.0(1)\%$, with $2\%$ of the error due to off-resonant transitions to the adjacent $X^2\Sigma (v=0,N=1, J=3/2, F=2, m_F=1)$ state. These can be improved with pulse-shaping. The residual inefficiency is likely due to decoherence arising from differential ac Stark shifts. The microwave transfer efficiency from $\left|+\ket$ to $\left |{\uparrow}\ket$ is measured to be $99.1(1)\%$. The inefficiency is again due to off-resonant transfers and decoherence due to differential ac Stark shifts. We find that the resonant clean-out pulse has a negligible effect on $|{+}\rangle$ molecules ($\sim0.1\%$ loss).

Overall, the fidelity of preparing single molecules in $\left|\uparrow\ket$ is 82.4(11)\%, with the primary error being empty tweezers. We estimate that $0.9(1)\%$ of molecules remain in $\left|+\ket$ due to imperfections in the final microwave sweep. Both of these errors are state preparation errors, since the system is prepared outside the Hilbert space spanned by $\{\left|\uparrow \ket, \left|\downarrow\ket \}$. These errors can be corrected for when estimating the state preparation and measurement (SPAM) corrected Bell state fidelity, as described in detail in a subsequent section.

\subsection{XY8 Decoupling Sequence}
For the $P_{11}$ oscillation data presented in Fig.\,3(d,e,f) in the main text, XY8 dynamical decoupling pulses are applied during the free evolution time. The specific XY8 parameters used are as follows. 10.6\,$\mu\text{s}$ square pulses are used to implement the $\pi$ pulses. For tweezer separations of $1.26(2)\,\mu\text{m}$ to $1.59(2)\,\mu\text{m}$, which correspond to AOD frequency separations of 300\,kHz to 380\,kHz, $\tau=100\,\mu\text{s}$ and each XY8 block is repeated every 1.6\,ms. Due to memory limitations on our experimental control system, for data at the larger separations of $1.68(2)\,\mu\text{m}$ to $2.35(3)\,\mu\text{m}$ (AOD frequency separations of 400\,kHz to 560\,kHz), $\tau=200\,\mu\text{s}$, and each XY8 block is repeated every 3.2\,ms. The reader can refer to Fig. 3(b) for a diagram defining $\tau$ in the XY8 sequence.

\subsection{Extracting $J$ From $P_{11}$ Oscillations}
The data shown in Fig.~3(d,e,f) in the main text is fit to a phenomenological model, in order to extract the interaction strength $J$. The model has doubly damped oscillations, with one damping parameter capturing population loss, and the second capturing additional damping of the oscillations:
\begin{equation}
P_{\uparrow\uparrow} = A e^{-\gamma_1 t} \left(1-\exp(-\gamma_2 t) \cos \left(\frac{J t}{2}\right)\right).
\end{equation}

\subsection{Modeling Finite Temperature Effects on $P_{11}$ Oscillations }
\label{sec:thermal}
To model the finite motional temperature of molecules in the trap, we begin by approximating each tweezer trap by a 3D harmonic oscillator with a trapping potential described by $V(\vec{r}) = \frac{1}{2}m (\omega_r^2 (x^2+y^2) + \omega_z^2 z^2)$, where $\omega_z$ and $\omega_r$ are measured directly via modulation spectroscopy~\cite{holland2022bichromatic}. We take a pair of tweezers to be separated by a distance $d$ along $\hat{x}$. For two molecules positioned at $\vec{r}_1$ and $\vec{r}_2$, we define the center-of-mass coordinate $\vec{R} = \frac{1}{2} (\vec{r}_1+\vec{r}_2)$ and relative coordinate $\vec{r} = \vec{r}_2-\vec{r}_1-d\hat{x}$. The resulting Hamiltonian can be written as
\begin{eqnarray}
\hat{H} &=& \hat{H}_{\text{CM}} + \hat{H}_\text{r} \nonumber \\
\hat{H}_{\text{CM}} &=& -\frac{\hbar^2}{2M} \nabla_R^2 + \frac{1}{2} M (\omega_r^2 (X^2+Y^2) + \omega_z^2 Z^2) \nonumber \\
\hat{H}_{\text{r}} &=& - \frac{\hbar^2}{2\mu} \nabla_r^2 + \frac{1}{2} \mu (\omega_r^2 (x^2+y^2) + \omega_z^2 z^2) + \hat{U}(\vec{r})\nonumber \\
\hat{U} &=&  \frac{C}{|\vec{r} + d\, \hat{x}|^3}\left(1-\frac{3y^2}{|\vec{r} + d\, \hat{x}|^2}\right) (\left| \uparrow \downarrow \ket \bra \downarrow \uparrow \right| + c.c.)\nonumber,
\end{eqnarray}
where $M=2m$ and $\mu=m/2$, and 
\begin{eqnarray}
\label{eq:C}
C &=& \frac{1}{4\pi \varepsilon_0}\bra \uparrow \downarrow\right| \left[T^1_0(\mathbf{d}_1) T^1_0(\mathbf{d}_2) \right.\nonumber \\
& &\left.+\left(T^1_1(\mathbf{d}_1) T^1_{-1}(\mathbf{d}_2)+T^1_{-1}(\mathbf{d}_1) T^1_{1}(\mathbf{d}_2)\right)\right]\left|\downarrow \uparrow\ket,
\end{eqnarray}
with $T^p_q(\mathbf{A})$ denoting the $q^{\text{th}}$ component of the rank-$p$ spherical tensor representation of an operator $\mathbf{A}$.

For our choice of interacting states $\left|\uparrow\ket = \left|X^2\Sigma(v=0), N=1,J=1/2,F=0,m_F=0\ket$ and  $\left|\downarrow\ket = \left|X^2\Sigma(v=0), N=0,J=1/2,F=1,m_F=0\ket$, only the first term in Eq.~\ref{eq:C} is non-zero, giving
\begin{equation}
C = \frac{d^2}{4\pi \varepsilon_0},
\end{equation}
where $d = -0.328(8)\, d_X$, $d_X=-1.21(3 )\,e a_0$ being the $X$-state dipole moment of CaF~\cite{Childs1984CaFXd}. Note that since we make use of a $\Delta m_F=0$ transition, $C$ is positive and the resulting spin-exchange interactions are anti-ferromagnetic. This is as expected as dipoles aligned side-by-side experience repulsive interactions. Our case is in contrast to those in~\cite{yan2013observation,christakis2022probing}, where $|\Delta m_F|=1$ transitions are used, and $C$ is negative, leading to ferromagnetic interactions ($T^1_{1}(\mathbf{d})= -T^1_{-1}(\mathbf{d})^*$). The sign change in interactions between using $\Delta m_F=0$ and $|\Delta m_F|=1$ transitions was recently exploited in~\cite{li2022tunable}.

For the interaction operator $\hat{U}$, we approximate the position operators $x,y,z$ as classical parameters. We next expand the interaction $\hat{U}$ to second order in $x$, $y$ and $z$ to obtain
\begin{eqnarray}
\hat{U} &=& \eta J_0  \left[\left| \uparrow \downarrow \ket \bra \downarrow \uparrow \right| + c.c.\right]\nonumber \\
\eta &=& \left(1 -\frac{3x}{d}+  \frac{6x^2}{d^2} - \frac{9y^2}{2d^2} -\frac{3 z^2}{2d^2} \right), \nonumber
\end{eqnarray}
where $J_0 = C/d^3$ is the interaction strength for two point particles separated by $d$, and $\eta$ is the correction factor due to thermal motion. In our system, the tweezer trap period is much shorter than the interaction timescale (ie, $\omega_r,\omega_z \gg J$). This means that an effect analogous to motional narrowing should occur generically and suppress the variation in $\eta$ due to terms linear in $x$. We note that the agreement between the observations derived using this assumption indirectly suggests that this is indeed true, but direct confirmation of this effect is left for future work. The suppression of variation in $\eta$ due to the linear term can also be derived using a quantum mechanical approach.

In the absence of motion and decoherence, $P_{11}$ oscillates with angular frequency $J/2$. To capture the effects of finite temperature, we create multiple instances of $\eta$ for a thermal distribution of separations ${x,y,z}$, taking into account only terms quadratic in ${x,y,z}$. The thermal distribution used is $P(x,y,z) = P_x(x) P_y(y) P_z(z)$, where
\begin{eqnarray}
P_x (x) &=& \frac{1}{\mathcal{N}_x} \exp \left (-\frac{\mu \omega_r^2 x^2}{2k_B T}\right) \nonumber \\
P_y (y) &=& \frac{1}{\mathcal{N}_y} \exp \left (-\frac{\mu \omega_r^2 y^2}{2k_B T}\right)\nonumber \\
P_z (z) &=& \frac{1}{\mathcal{N}_z} \exp \left (-\frac{\mu \omega_z^2 z^2}{2k_B T}\right), \nonumber
\end{eqnarray}
and $\mathcal{N}_i$ are normalization constants such that $\int dx\, dy\, dz\, P(x,y,z) = 1$.

For each tweezer separation, we then average $\cos (\eta t)$ over all instances of $\eta$ drawn from the thermal distributions. The resulting average curve $\bra \cos (\eta t)\ket_\eta$ is then fit using a damped sinusoidal function $f(t) = A(1-e^{-\gamma t}\cos(\eta_{\text{eff}}\, t))$ over 2 cycles, approximately matching the time duration where oscillations are visible in the experiment. The fitted value of $\eta_{\text{eff}}$ gives the correction factor to $J_0$. This leads to oscillations in $P_{11}$ appearing at an effective angular frequency $J_{\text{eff}} = J_0\eta_{\text{eff}}$. As seen in Fig.~3f, the resulting estimates agree well with the observed $P_{11}$ oscillation frequencies.

\subsection{Damping of $P_{11}$ Oscillations }
In this section, we discuss in detail damping of $P_{11}$ oscillations. As described in the main text, we find that after correcting for single molecule loss, there is still damping of the $P_{11}$ curves. This can be due to many factors such as single particle decoherence and motional dephasing. We attempt to quantify these two causes.

We first develop a model for single particle dephasing. For simplicity, we assume a decoherence model of only bit flips and phase flips occurring at rates of $\gamma_B$ and $\gamma_P$. The density matrix $\rho$ of a single 2-level system evolves according to a master equation in Lindblad form:
\begin{equation}
\dot{\rho} = -\frac{i}{\hbar} \left[\hat{H}, \rho\right] + \sum_i \gamma_i \left(L_i \rho L_i^\dag -\frac{1}{2} \left\{L_i \rho L_i^\dag , \rho \right\}\right),
\end{equation}
with $L_1 = \sigma_x/\sqrt{2}$, $L_2=\sigma_z/\sqrt{2}$, $\gamma_1 = \gamma_B$ and $\gamma_2 = \gamma_P$. $\gamma_B$ and $\gamma_P$ are extracted from separate measurements of the decays of $\hat{S}_z$ and $\hat{S}_y$ at the various tweezer separations used and under identical dynamical decoupling parameters. In detail,  the decay of $\hat{S}_z$ is determined by measuring $P_{\uparrow}$ and $P_{\downarrow}$ after holding molecules initialized in $\left|\uparrow \ket$ for a variable time. The decay of $\hat{S}_y$ is determined by measuring the decay of Ramsey contrast. The decays are first corrected for the directly measured molecular loss before being fit to an exponential decay. The decay of $\hat{S}_z$ gives $\gamma_B$, and the decay of $\hat{S}_y$ gives $\gamma_B+ \gamma_P$. We verify that $\hat{S}_y$ indeed decays faster than $\hat{S}_z$.

Having obtained single particle decoherence parameters $\gamma_B$ and $\gamma_P$,  we assume that for two molecules, the density matrix still evolves according to a master equation in Lindblad form. We further assume that the system is described by four Lindblad operators given by $L_1 = \sigma_x \otimes \mathbf{1} /\sqrt{2}$ , $L_2 =  \mathbf{1} \otimes \sigma_1  /\sqrt{2}$, $L_3 = \sigma_z \otimes \mathbf{1} /\sqrt{2}$ and $L_4 =   \mathbf{1}\otimes \sigma_z /\sqrt{2}$, with corresponding rates of $\gamma_1 = \gamma_2=\gamma_{b,2}$ and $\gamma_3=\gamma_4= \gamma_{p,2}$. In the absence of loss, it turns out that under these assumptions, the damping of $P_{\uparrow\uparrow}$ is only determined by $\gamma_{\text{tot}}=\gamma_{b,2}+\gamma_{p,2}$, with $P_{\uparrow\uparrow}$ given by 
\begin{equation}
P_{\uparrow\uparrow} =\frac{1}{4}\left( 1+ e^{-2\gamma_{\text{tot}} t}-2 e^{-2\gamma_{\text{tot}} t} \cos\left(\frac{Jt}{2} \right)\right).
\label{eq:decohmodel}
\end{equation}

\begin{figure}
	{\includegraphics[width=\columnwidth]{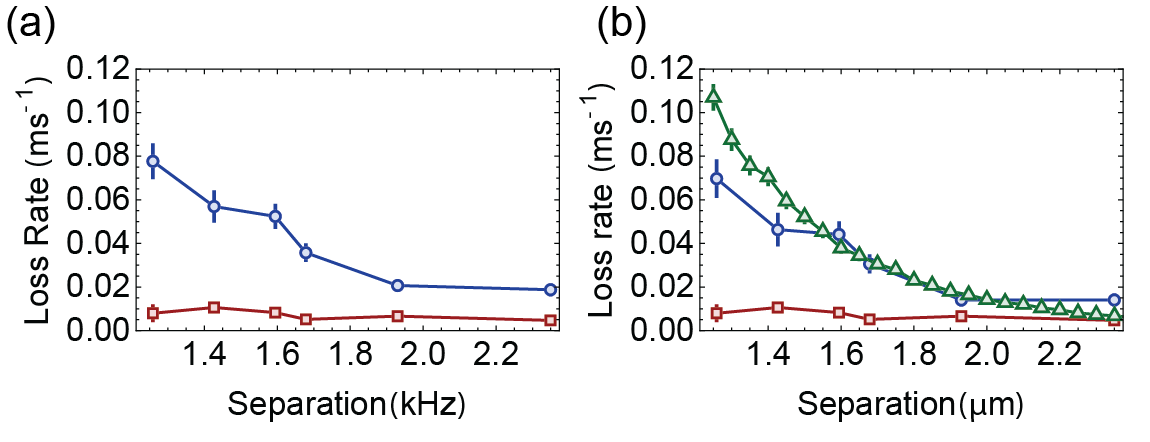}}
	\caption{\label{fig:P11decoh}Phenomenological $P_{\uparrow\uparrow}$ Oscillation Damping Rates (a) Shown in blue circles is the extracted effective damping rate $\gamma_{\text{tot}}$ for $P_{\uparrow\uparrow}$ oscillations, versus tweezer separation. Shown in red squares is the measured single particle decoherence rate $\gamma_{\text{tot},1}$. (b) Shown in blue circles is the extracted residual damping rate $\gamma_{r}$ for $P_{\uparrow\uparrow}$ oscillations, versus tweezer separation. Shown in green triangles squares is the effective damping rate $\gamma_{\text{therm}}$ due to thermal dephasing. Shown in red squares is the measured single particle decoherence rate $\gamma_{\text{tot},1}$ versus tweezer separation. }
	\vspace{-0.2in}
\end{figure}

We fit the experimental data to this model to obtain a phenomenological value of $\gamma_{\text{tot}}$. In detail, we first correct the experimental data by 1) the measured population decay and 2) the state preparation fidelity to obtain $P_{\uparrow\uparrow}'$. $P_{\uparrow\uparrow}'$ is then fit it to Eq.~(\ref{eq:decohmodel}) to obtain $\gamma_{\text{tot}}$. The resulting curves fit the data well. The extracted values of $\gamma_{\text{tot}}$ are higher at shorter distances. In Fig.~\ref{fig:P11decoh}(a), we show the fitted $\gamma_{\text{tot}}$ along with $\gamma_{\text{tot},1}=\gamma_B+\gamma_P$, the sum of the measured single particle decoherence $\gamma_B$ and $\gamma_P$. We observe that $\gamma_{\text{tot}}$ decrease with larger separations, but is higher than $\gamma_{\text{tot},1}$, indicating that additional damping effects are likely present. One possible cause is thermal dephasing. As discussed in the previous section, due to finite temperature, molecules interact with slightly different strengths for each experimental run. This can lead to effective dephasing of $P_{\uparrow\uparrow}$ oscillations.

In Fig.~\ref{fig:P11decoh}(b), we show the residual damping rate $\gamma_r = \gamma_{\text{tot}}- \gamma_{\text{tot},1}$. This reveals the measured excess damping rate after single particle decoherence (under our assumptions) is removed. For comparison, we show the expected effective damping rate $\gamma_{\text{therm}}$ due to thermal motion, obtained from the procedure in the previous section. We find that thermal motion predicts similar levels of damping compared to the data, and qualitatively reproduces the dependence on tweezer separation. 

We caution that the modeling of damping and decoherence presented here has several unverified assumptions, and future investigations are needed to fully understand the observed damping of $P_{\uparrow\uparrow}$. Nevertheless, we can conclude that additional damping unexplained by molecular loss is likely present, and thermal dephasing could be a possible explanation.

\subsection{Measuring the Bell State Fidelity}
\label{sec:bellfidelity}
We consider a Bell state of the form $\left|\psi_B\ket = \left(\left|\uparrow \uparrow\ket + e^{i \phi} \left|\downarrow\downarrow\ket\right)/\sqrt{2}$. The Bell state fidelity $\mathcal{F}$ for a system described by a density matrix $\rho$ is given by~\cite{turchette1998deterministic,Sackett2000GHZ,Levine2019gate,Madjarov2020entangle,Schine2022gate,Ma2022YbGate}
\begin{equation}
\mathcal{F} =\bra \psi_B | \rho| \psi_B\ket.
\end{equation}
It follows that the maximum fidelity $\mathcal{F}$ of creating a Bell state with some $\phi$ is given by
\begin{equation}
\mathcal{F} = \frac{1}{2} \left(P_{\uparrow\uparrow} + P_{\downarrow\downarrow} + \mathcal{C} \right)
\end{equation}
where $P_{\sigma \sigma} =\rho_{\sigma \sigma \sigma \sigma}$ are the populations in the states $\left |\sigma \sigma \ket$, and $\mathcal{C} = 2|\rho_{\uparrow \uparrow \downarrow\downarrow}|$ is proportional to the magnitude of the two-particle coherence $\rho_{\uparrow \uparrow \downarrow\downarrow}$. 

For a 2-level system, the two-particle coherence $\mathcal{C}$ can be measured via parity oscillations~\cite{turchette1998deterministic}. In detail, one applies a global $\pi/2$ pulse along a variable rotation axis $\hat{n} = \cos \theta\,\hat{x} + \sin\theta\,\hat{y}$, and measures the resulting parity signal given by
\begin{eqnarray}
\bra \hat{\Pi}^{(2)} \ket &=& \bra  \hat{\sigma}_z \otimes \hat{\sigma}_z \ket \nonumber \\
&= &\bra \left(\left|\uparrow \uparrow \ket\bra \uparrow \uparrow \right|  + \left|\downarrow \downarrow \ket\bra \downarrow \downarrow \right| - \left|\uparrow \downarrow \ket\bra \uparrow \downarrow \right| - \left|\downarrow \uparrow \ket\bra \downarrow \uparrow  \right| \right) \ket \nonumber \\
&=& P_{\uparrow\uparrow} +P_{\downarrow\downarrow} -P_{\uparrow\downarrow} -P_{\downarrow\uparrow}, \nonumber
\end{eqnarray} 
For a general density matrix $\rho$, this gives
\begin{equation}
 \bra \hat{\Pi}^{(2)} \ket =2|\rho_{\uparrow\downarrow \downarrow \uparrow } |\cos (\varphi_{\uparrow\downarrow \downarrow \uparrow })+  2|\rho_{\uparrow\uparrow\downarrow\downarrow}| \cos(2\theta - \varphi_{\uparrow\uparrow\downarrow\downarrow}) ,
\end{equation}
where $\varphi_i = \text{Arg} [\rho_i$]. 

In our experiment, we only detect molecules in $X^2\Sigma(v=0,N=1)$. This means that tweezers filled with $\left|\uparrow\ket$ molecules are directly detected as 1s, while tweezers filled with $\left|\downarrow\ket$ molecules and empty tweezers appear as 0s. In addition, imperfect internal state preparation leaves molecules in the $\left |+\ket$ state, which also appear dark. Tracing over all states where the tweezer is not occupied by a  $\left|\uparrow \ket$ or $\left| \downarrow\ket $ molecule, the two particle Hilbert space is spanned by $\{\uparrow, \downarrow, e\} \otimes \{\uparrow, \downarrow, e\}$, with $\left| e \ket$ denoting the state where the tweezer is neither occupied by a $\left|\uparrow \ket$ nor $\left| \downarrow\ket $ molecule.

In this basis, the single particle observable that can be directly measured is given by
\begin{equation}
\sigma_z^{(3)} = \left(
\begin{array}{ccc} 
1 & 0 & 0 \\ 
0 & -1 & 0 \\ 
 0 & 0 & -1 
 \end{array}
\right),
\end{equation}
which does not distinguish between empty tweezers and tweezers with molecules in $\left|\downarrow\ket$, $\left|+\ket$.

Rather than directly measuring $\hat{\Pi}^{(2)}$, we measure the effective parity $\bra \hat{\Pi} \ket = \bra \sigma_z^{(3)} \otimes \sigma_z^{(3)} \ket = P_{11} + P_{00} - P_{10} - P_{01}$, where the $P_{ij}$ denote probabilities, and 1(0) denotes bright (dark) tweezers. One can show that the resulting effective parity is in general a sum of sinusoidal oscillations periodic in $\theta$ and $2\theta$. The amplitudes of the oscillations periodic in $\theta$ are related to single particle coherences such as $\rho_{e \uparrow,e \downarrow}$, while the amplitude of the $2\theta$ oscillation is $2|\rho_{\uparrow\uparrow\downarrow\downarrow}| = \mathcal{C}$. In detail, for a general density matrix $\rho$ in the space spanned by $\{\left|\uparrow\ket, \left| \downarrow\ket, \left|e\ket\} \otimes\{\left|\uparrow\ket, \left| \downarrow\ket, \left|e\ket\}$, the effective parity signal following $\hat{R}_\theta(\pi/2)$ is given by
\begin{eqnarray}
 \bra \hat{\Pi} \ket &=& \rho_{e e e e } +  2|\rho_{\uparrow\downarrow \downarrow \uparrow } |\cos (\varphi_{\uparrow\downarrow \downarrow \uparrow })+ \nonumber \\
 & &  2|\rho_{e \uparrow e \downarrow}|  \cos (\theta- \varphi_{e \uparrow e \downarrow})+  2|\rho_{\uparrow e \downarrow e }| \cos (\theta- \varphi_{\uparrow e \downarrow e })   \nonumber \\
 & & + 2|\rho_{\uparrow\uparrow\downarrow\downarrow}| \cos(2\theta - \varphi_{\uparrow\uparrow\downarrow\downarrow}) .
\label{eq:pigeneral}
\end{eqnarray}

\subsection{State Preparation Correction}
\label{sec:statepreperr}
In this section, we show how to correct the raw measurement-corrected Bell fidelity $\mathcal{F}$ by imperfect state preparation. Imperfect state preparation includes the primary error of producing empty tweezers, and the small error of producing molecules in the state $\left|+\ket$. As we will show, the SPAM-corrected fidelity $\mathcal{F}_{\text{SPAM}}$ can be obtained from the measurement-corrected raw fidelity $\mathcal{F}$ via
\begin{equation}
\mathcal{F}_{\text{SPAM}} = \mathcal{F}/p_2,
\end{equation}
where $p_2$ is the probability of initializing the two-tweezer system in the subspace spanned by $\{\left|\uparrow\ket,\left|\downarrow\ket\}\otimes\{ \left|\uparrow\ket,\left|\downarrow\ket\}$. To obtain $p_2$, we first apply the state preparation procedure to prepare molecules in $\left|\uparrow\ket$ and detect the fraction of molecules $p_1$ that are indeed in $\left|\uparrow\ket$. $p_2$ is then found via $p_2=p_1^2$.

To show this, we perform a partial trace over states outside of the relevant subspace of $\{  \{\left| \uparrow\ket, \left |\downarrow\ket\}  \otimes \{\left| \uparrow\ket, \left |\downarrow\ket\},\left|\emptyset\ket \}$, where $\left|\emptyset\ket$ denotes the state where at least one tweezer is empty or has a molecule not initialized in $\left| \uparrow\ket$ or $\left |\downarrow\ket$. Concretely, this can be done by expanding the Hilbert space through second-quantization.

In detail, each tweezer has $3+n$ modes that can be empty or filled. The first 3 modes correspond to $\left|\uparrow\ket$, $\left|\downarrow\ket$ and an empty tweezer respectively, while the remaining $n$ modes correspond to single particle molecular states outside of $\left|\uparrow\ket$ and $\left|\downarrow\ket$. The physical states are those with one total excitation over all the modes. 
The Hilbert space for two tweezers is then spanned by $(2^{3+n})^2$ states, of which $(3+n)^2$ are physical.

For example, in our system, the physical states are $\{\left |\uparrow\ket, \left |\downarrow\ket, \left |e\ket,\left |+\ket\} \otimes \{\left |\uparrow\ket, \left |\downarrow\ket, \left |e\ket,\left |+\ket\}$, where $\left|e\ket$ denotes the state with an empty tweezer. In other words, $n=1$ with the extra mode being $\left| +\ket$. The expanded Hilbert space allows us to use basis states in product-state form, and trace over mode other than the first two. This correspond to tracing out states where at least one tweezer is in a mode outside of $\left|\uparrow\ket$ or $\left|\downarrow\ket$. For the density matrix in the expanded Hilbert space, populations and coherences involving unphysical states can be set to zero when performing the partial trace over the $n+1$ modes.

One can show that after tracing over the modes corresponding to $\left|+\ket$ and $\left|e\ket$, the reduced density matrix $\rho^{\text{red}}$ is of block-diagonal form and has dimensions of $2^{4}\times 2^4$. Restricting to the 5 physical states $ \{\left| \uparrow \uparrow\ket, \left |\uparrow\downarrow\ket \left| \uparrow\downarrow\ket, \left |\downarrow\downarrow\ket, \left|\emptyset\ket \}$, the reduced density matrix can be written as 
\begin{equation} 
\rho_{\text{eff}} = \rho_2 \oplus \rho_\emptyset,
\label{eq:eff}
\end{equation}
where $\rho_2$ is spanned by $\left| \uparrow \uparrow\ket, \left |\uparrow\downarrow\ket \left| \uparrow\downarrow\ket, \left |\downarrow\downarrow\ket$, and $\rho_\emptyset$ is spanned by $\left|\emptyset\ket\bra \emptyset \right|$. $\rho_\emptyset$ has one entry $\rho_{\emptyset\emptyset}$, which is precisely the probability $1-p_2$ of obtaining at least one tweezer outside of  $\{\left| \uparrow\ket, \left |\downarrow\ket\}$. Imperfect state preparation (including preparing empty tweezers and molecules in the wrong internal state) can therefore simply be described scaling the 4x4 block $\rho_2$ by $p_2$. The SPAM-corrected density matrix $\rho_{\text{SPAM}}$ for the ideal two spin-1/2 system can therefore be obtained from $\rho_2$ by
\begin{equation}
\rho_{\text{SPAM}} = \frac{\rho_2}{\text{Tr}[\rho_2]} = \frac{1}{p_2} \rho_2
\end{equation}
Since the raw fidelity $\mathcal{F}$  (measurement-corrected) is obtained as a sum of various terms in $\rho_2$, it follows that the SPAM-corrected fidelity $\mathcal{F}_{\text{SPAM}}$ is given by $\mathcal{F}_{\text{SPAM}} = \mathcal{F}/p_2$. The factor $p_2$ is directly measured in the experiment.

For concreteness, we show a calculation with $n=0$. In this case, there are 9 physical two-tweezer states spanned by $\{\left |\uparrow\ket, \left |\downarrow\ket, \left |e\ket\} \otimes \{\left |\uparrow\ket, \left |\downarrow\ket,\left |e\ket\}$. In the expanded Hilbert space, each basis state can be described by two strings with 1s and 0s, each with a length of $3$. Physical states are those for which both strings have exactly a single 1. For example, the state $\left|100;001\ket$ is physical and corresponds to $\left |\uparrow\ket \otimes \left |e\ket$. We next obtain the reduced density matrix after tracing over the mode corresponding to $\left|e\ket$ for each of the tweezers.

Suppose we want to find the entry  $\rho^{\text{red}}_{\uparrow\uparrow\downarrow\downarrow}$ in the reduced density matrix $\rho^{\text{red}}$. This is given by
\begin{equation}
\rho^{\text{red}}_{\uparrow\uparrow\downarrow\downarrow} = \sum_{i,j=0} ^{1} \rho_{10i;10j,01i;01j}
\end{equation}
The only physical term is one where $i=j=0$, as physical states must only contain a single 1 in each 3-string. Since non-physical entries in $\rho$ are zero, only a single term corresponding to $\rho_{\uparrow\uparrow\downarrow\downarrow}$ contributes, as expected.

Now suppose we want to find the off-diagonal entry $\rho^{\text{red}}_{\uparrow\uparrow\emptyset\downarrow}$ in the reduced density matrix, where $\left |\emptyset\ket$ denotes the state when the tweezer has neither a $\left|\uparrow\ket$ or $\left|\downarrow\ket$ molecule. One finds that this is given by
\begin{equation}
\rho^{\text{red}}_{\uparrow\uparrow\emptyset\downarrow} = \sum_{i,j=0} ^{1} \rho_{10i;10j,00i;01j}
\end{equation}
Since the only physical states are ones where each 3-string has only a single 1, one sees that no density matrix elements corresponding to physical states contribute. Consequently, $\rho_{2,\uparrow\uparrow\emptyset\downarrow} =0$. In general, one can show that there are no remaining coherences between $\left|\emptyset\ket$ and  $\left|\uparrow\ket$ or $\left|\downarrow\ket$. Therefore, $\rho^{\text{red}}$ is block-diagonal. Removing unphysical states that only contribute to vanishing entries in $\rho^{\text{red}}$, one obtains $\rho_{\text{eff}}$, the 5x5 reduced density matrix given by Eq.~\ref{eq:eff}. The subsequent argument presented above for obtaining $\mathcal{F}_{\text{SPAM}}$ from $\rho_2$ can then be used.

\subsection{Extracting Full Population Distributions}
Since we only detect population in $X^2\Sigma(v=0,N=1)$, occupied tweezers with molecules in $|{\uparrow}\rangle$ appear bright (1), whereas occupied tweezers in $|{\downarrow}\rangle$ or empty tweezers $|{e}\rangle$ appear dark (0). To obtain the Bell state fidelity, $P_{\uparrow\uparrow}$ and $P_{\downarrow\downarrow}$ are both needed. The former directly coincides with $P_{00}$ while the latter is obtained by applying an additional $\pi$ pulse prior to detection. 

Under certain assumptions, which we describe in detail here, the full populations can be obtained from the two measurements, which we denote $\{P_{11},P_{10},P_{01},P_{00}\}$ and $\{\bar{P}_{11},\bar{P}_{10},\bar{P}_{01},\bar{P}_{00}\}$ with and without the $\pi$-pulse, respectively. 

The assumption needed to obtain the full distribution is that the two tweezers comprising each tweezer pair are identical. It follows that $P_{10}= P_{01}$ and $\bar{P}_{10}=\bar{P}_{01}$. These two conditions are directly verified by our measurements. Instead of 8 measured quantities, the symmetrization assumption reduces the measured quantities to 6:
\begin{eqnarray}
p_{ii} &=& P_{ii}\nonumber \\
\bar{p}_{ii} &=& \bar{P}_{ii}\nonumber \\
p_{ij} &=& P_{ij} + P_{ji}, j>i \nonumber \\
\bar{p}_{ij} &=& \bar{P}_{ij} + \bar{P}_{ji}, j>i. \nonumber 
\end{eqnarray}
Making use of the normalization conditions $p_{11}+p_{10}+p_{00}=1$, $\bar{p}_{11}+\bar{p}_{10}+\bar{p}_{00}=1$, there are only 4 remaining degrees of freedom. We can encode these in a single vector $\mathbf{p}_M$ as
\begin{equation}
\mathbf{p}_M=
\begin{pmatrix}
p_{11} & p_{10}  & \bar{p}_{11} & \bar{p}_{10}
\end{pmatrix}.
\end{equation}

Next we consider the full population distribution $\{P_{\uparrow\uparrow},P_{\uparrow\downarrow},P_{\downarrow\uparrow},P_{\downarrow\downarrow},P_{\uparrow e},P_{e\uparrow},P_{\downarrow e},P_{e\downarrow},P_{e e} \}$, where $\left|e\ket$ denotes the tweezer state where the tweezer is either empty or occupied by a molecule not in $\left|\uparrow\ket$ or $\left|\downarrow\ket$. The symmetry of the two tweezers leaves 6 independent probabilities $\{p_{\uparrow\uparrow}, p_{\uparrow\downarrow}, p_{\downarrow\downarrow}, p_{\uparrow e}, p_{\downarrow e} ,p_{ee}\}$, where $2p_{\sigma e}  = P_{\sigma e} = P_{e\sigma}$, $\sigma = \uparrow, \downarrow$, $p_{\sigma\sigma'} = P_{\sigma\sigma'}$ and $p_{ee} = P_{ee}$. Using the normalization condition $p_{\uparrow\uparrow} + p_{\uparrow\downarrow} + p_{\downarrow\downarrow} + p_{\uparrow e} + p_{\downarrow e} + p_{ee}=1$, 5 independent probabilities are left. These are captured by a single vector $\mathbf{p}_T$:
\begin{equation}
\mathbf{p}_T=
\begin{pmatrix}
p_{\uparrow\uparrow} & p_{\uparrow\downarrow} & p_{\downarrow\downarrow} & p_{\uparrow e} & p_{\downarrow e}
\end{pmatrix}.
\end{equation}
\noindent

The measurement procedure is a linear transformation between the probabilities $\mathbf{p}_T$ and the measured probabilities $\mathbf{p}_M$. This can be expressed as $\mathbf{p}_M=\mathbf{T}.\mathbf{p}_T$, where
\begin{equation}
\mathbf{T}=
\begin{pmatrix}
1 & 0 & 0 & 0 & 0  \\
0 & 1 & 0 & 1 & 0  \\
0 & 0 & 1 & 0 & 0  \\
0 & 1 & 0 & 0 & 1 \\
\end{pmatrix}
\end{equation}
The null space of $\mathbf{T}$ has dimension one, and is spanned by
\begin{equation}
\mathbf{p}_N =
\begin{pmatrix}
0 & 1 & 0 & -1 & -1 
\end{pmatrix}.
\end{equation}
This indicates that if $p_{\uparrow\downarrow}$ is increased by the same amount as $p_{\uparrow e}$ and $p_{\downarrow e}$ are decreased, there will be no observable difference in the measured probabilities $\mathbf{p}_M$. The system is under-constrained by one degree of freedom. 

By introducing an additional constraint from state preparation, the system can in fact be fully constrained. To fully constrain the system, we impose that the probabilities of doubly-occupied, singly-occupied, and empty tweezer pairs agree with an effective single-particle state preparation fidelity $f$ (after generating the Bell state). Explicitly, this constraint is

\begin{align}\label{eq:singStatePrepConst}
f^2 &= p_{\uparrow\uparrow} + p_{\uparrow\downarrow} + p_{\downarrow\downarrow} \\\nonumber
2f(1-f)&= p_{\uparrow e} + p_{\downarrow e} \\\nonumber
(1-f)^2 &= p_{ee}.
\end{align}
Note that the last equation follows from the first two equations under the normalization condition for $\mathbf{p}_T$. This nonlinear constraint allows $\mathbf{p}_T$ to be fully determined.

One can verify that 
\begin{eqnarray}
p_{\uparrow\uparrow}&=&p_{11} \nonumber \\
p_{\downarrow\downarrow}&=&\bar{p}_{11} \nonumber \\
p_{\uparrow\downarrow} &=& \frac{1}{2}(p_{10}+\bar{p}_{10}-p_\text{sum})\nonumber \\
p_{\uparrow e} &=& \frac{1}{2}(p_\text{sum}-p_\text{diff}) \nonumber \\
p_{\downarrow e} &=& \frac{1}{2}(p_\text{sum}+p_\text{diff}), \nonumber
\end{eqnarray}
where
\begin{eqnarray}
p_\text{sum} & =& \left(1-\frac{p_{10}+\bar{p}_{10}}{2} - (p_{11}+\bar{p}_{11})\right)^2 \nonumber \\
p_\text{diff} &=& \bar{p}_{10}-p_{10} \nonumber \\
\end{eqnarray}
Alternative expressions for $p_\text{sum}$ and $p_\text{diff}$ are
\begin{eqnarray}
p_\text{sum} &=& \left(\frac{a+b}{2}\right)^2 \nonumber \\
p_\text{diff} &=& a-b  \nonumber \\
a&=& p_{00} - \bar{p}_{11} \nonumber \\
b&=& \bar{p}_{00} - p_{11}, \nonumber 
\end{eqnarray}
which are equivalent to the first expressions because of normalization conditions.

\begin{figure}
	{\includegraphics[width=\columnwidth]{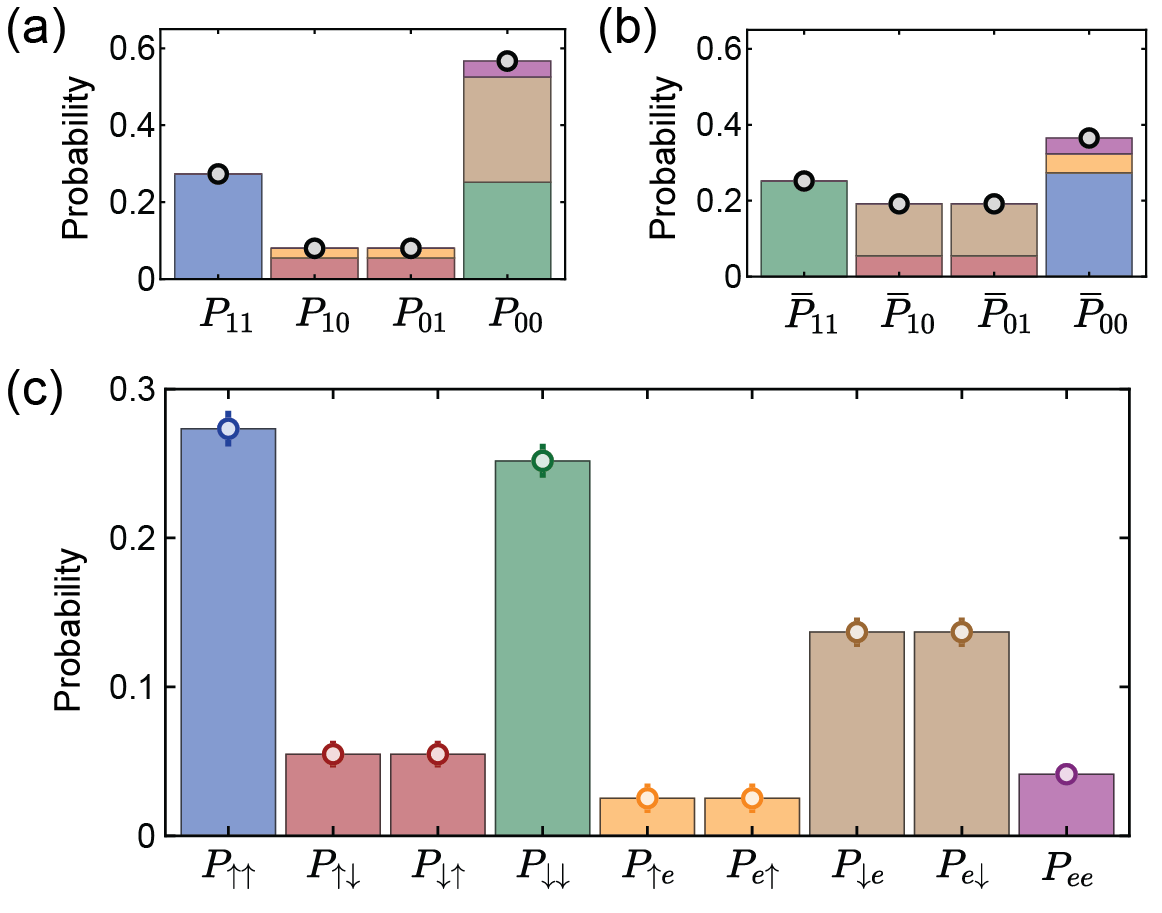}}
	\caption{\label{fig_9barcolors} Two-tweezer probabilities (a,b) The reconstructed measurement outcomes without (a) and with (b) a $\pi$ pulse prior to detection. The various colors show contributions from the full probabilities shown in (c). (c) The full reconstructed probabilities. }
\end{figure}

In Fig.~\ref{fig_9barcolors}, we show the contributions to the reconstructed fluorescence histograms due to the true distributions extracted using the above procedure. Explicitly, we find

\begin{equation}
\begin{pmatrix}
p_{\uparrow\uparrow}\\
p_{\uparrow\downarrow}\\
p_{\downarrow\downarrow}\\
p_{\uparrow e}\\
p_{\downarrow e}\\
p_{ee}
\end{pmatrix}
=
\begin{pmatrix}
0.273(10)\\
0.110(14)\\
0.252(10)\\
0.050(16)\\
0.274(16)\\
0.041(4)
\end{pmatrix}\nonumber
\end{equation}
\noindent
and
\begin{equation}
f=0.797(11)\nonumber
\end{equation}

The obtained value of single-particle preparation fidelity $f$ differs slightly from the directly measured value of $p_{1}=82.4(11)\%$. In general, $p_{1}$ is an upper bound for $f$, since additional molecular loss and loss out of the $\left|{\uparrow}\ket$-$\left|{\downarrow}\ket$ can occur during the Bell state generation procedure. This is due to heating loss and off-resonant excitation to adjacent $m_F$ levels from the dynamical decoupling pulses. We estimate from separate population leakage measurements that these processes contribute $\approx 0.5\%$ of loss. The remaining small difference between $f$ and $p_{1}$ is likely due to instability in the polarization of the optical pumping beams over the course of the experiment, which can lead to degraded optical pumping efficiency.

\subsection{Relating Parity Oscillations to Increased Metrological Sensitivity}
The Bell state $\left|\psi_B\ket = \left(\left|\uparrow \uparrow\ket + e^{i \phi} \left|\downarrow\downarrow\ket\right)/\sqrt{2}$ is a maximally entangled two-particle state that is a realization of a GHZ state with $N=2$. GHZ states are states of the form $(\left|\downarrow\ldots \downarrow\ket + \left|\uparrow\ldots \uparrow \ket)/\sqrt{2}$. They are metrologically useful maximally entangled states that allow phase estimation at the Heisenberg limit. Under time evolution by a global operator $\hbar \delta \hat{S}_z$ of time $t$, the $\left| \uparrow \uparrow \ldots \uparrow \ket$ component acquires a phase of $N \delta t$, which is $N$ times faster than that of the single particle state $(\left|\downarrow\ket + \left|\uparrow \ket)/\sqrt{2}$. The phase uncertainty for the same measurement time is therefore improved by $\sqrt{N}$ compared to the standard quantum limit, leading to a metrological gain of $N$.

Here, we show that the $2\theta$ oscillations in $\bra \hat{\Pi} \ket$ directly reveals the faster phase accumulation of the Bell state ($N=2$ GHZ state) and hence can be interpreted as direct evidence of metrological advantage. First, we denote the action of a global $\pi/2$ pulse along a variable rotation axis $\hat{n} = \cos \theta\,\hat{x} + \sin\theta\,\hat{y}$ by the operator $\hat{R}_\theta(\pi/2)$. This operator can be written as $\hat{R}_\theta(\pi/2) = \hat{R}_z(-\theta) \hat{R}_x(\pi/2) \hat{R}_z(\theta)$, where $\hat{R}_x$ ($\hat{R}_z$) denote rotations about the $\hat{x}$ ($\hat{z}$) axis. We next note that  $R_z(\theta) = e^{-i \theta S_z}$, and that $R_z(-\theta)$ commutes with the parity operator $\hat{\Pi}$. Therefore, the observed parity is equivalent to measuring the parity of $\hat{R}_x(\pi/2) \hat{R}_z(\theta)\left|\psi\ket =\hat{R}_x(\pi/2) (e^{-i \theta S_z}\left|\psi\ket)$ for a general state $\left|\psi\ket$. This means that varying $\theta$ and measuring $\hat{\Pi}$ is equivalent to measuring $\hat{\Pi}$ after subjecting $\left|\psi\ket$ to time evolution by a global field of $-\hbar \theta \hat{S}_z/t$ for a time $t$. The sensitivity of $\bra \hat{\Pi} \ket$ to $\theta$ therefore can reveal the phase sensitivity of the system. 

We next explicitly work out the dependence for a single particle state and the Bell state. To simplify the expressions, without loss of generality, we rotate about $x$ rather than $y$ for the $\pi/2$ pulse. Suppose the system is prepared in the unentangled state $\left|\psi_1\ket = \left(\left|\uparrow e\ket + e^{i \phi} \left|\downarrow e\ket\right)/\sqrt{2}$. After application of $\hat{R}_y(\pi/2) \hat{R}_z(\theta)$, we obtain
\begin{eqnarray}
& & \hat{R}_x(\pi/2) \hat{R}_z(\theta)\left|\psi_1\ket \nonumber \\
&=& \hat{R}_x(\pi/2)  \frac{1}{\sqrt{2}} \left(e^{-i\theta/2}\left|\uparrow e\ket + e^{i\theta/2}\left|\downarrow e\ket\right) \nonumber \\
&=& -i \sin \left(\frac{\theta}{2} \right)\left|\uparrow e\ket + \cos  \left(\frac{\theta}{2}\right) \left|\downarrow e\ket.
\end{eqnarray}
Noting that $\left|\uparrow e\ket$($\left|\downarrow e\ket$) has negative (positive) parity, we find that $\bra \hat{\Pi}\ket$ is given by $\frac{1}{4}\cos \theta$, which has $2\pi$ periodicity in $\theta$.

Now suppose the system is prepared in the Bell state $\left|\psi_2\ket = \left(\left|\uparrow \uparrow \ket + e^{i \phi} \left|\downarrow \downarrow \ket\right)/\sqrt{2}$
After application of $\hat{R}_y(\pi/2) \hat{R}_z(\theta)$, we obtain
\begin{eqnarray}
& & \hat{R}_x(\pi/2) \hat{R}_z(\theta)\left|\psi_2\ket \nonumber \\
&=& \hat{R}_x(\pi/2)  \frac{1}{\sqrt{2}} \left(e^{-i\theta}\left|\uparrow \uparrow \ket + e^{i\theta}\left|\downarrow \downarrow\ket\right) \nonumber \\
&=& \frac{1}{\sqrt{2}}\left[\cos \theta \left(\left|\uparrow \uparrow \ket +\left|\downarrow \downarrow \ket \right)- i \sin \theta \left(\left|\uparrow \downarrow \ket +\left|\downarrow \uparrow \ket \right)\right]
\end{eqnarray}
Noting that $\left|\uparrow \uparrow\ket$($\left|\uparrow\downarrow \ket$) and $\left|\downarrow \downarrow \ket$($\left|\downarrow \uparrow\ket$) have positive(negative) parity, we find that $\bra \hat{\Pi}\ket$ is given by $\frac{1}{4}\cos(2 \theta)$, which has $\pi$ periodicity in $\theta$.

Our measurements (Fig.~4b) show that $\bra \hat{\Pi}\ket$ is predominantly modulated with a period of $\pi$. This indicates that the state we create with the Bell sequence indeed has twice the sensitivity to $\hat{S}_z$ fields compared to single particle states.

\subsection{Verifying Anti-Ferromagnetic Interactions}
From Eq.~\ref{eq:pigeneral}, we note that the starting phase of $2\theta$-periodic oscillations in parity $\Pi$ directly gives $\varphi_{\uparrow\uparrow\downarrow\downarrow}$. For a Bell state $\left |\psi_B\ket= \left(\left|\uparrow \uparrow\ket + e^{i \phi} \left|\downarrow\downarrow\ket\right)/\sqrt{2}$, $\varphi_{\uparrow\uparrow \downarrow \downarrow} = \phi$. In other words, the relative phase parameter $\phi$ of the Bell state $\left |\psi_B\ket$ is directly given by the phase $\varphi_{\uparrow\uparrow\downarrow\downarrow}$ of the parity oscillation signal. For our Bell state generation pulse sequence, $\phi= \text{Sgn}[J]\pi/2$. Examining the observed oscillation signal in Fig.~3b in the main text verifies that $J>0$, that is, the spin-exchange interactions are anti-ferromagnetic.